\documentclass[
        fleqn,
        usenatbib,
    ]{mnras}
    \usepackage{
        amsmath,
        amssymb,
        newtxtext,
        newtxmath,
        ae,
        aecompl,
        graphicx,
        booktabs,
        xcolor,
        soul,
    }
    \usepackage[T1]{fontenc}
    \usepackage[
        labelfont=bf, 
        font=scriptsize 
    ]{caption}
    \usepackage[caption=false]{subfig} 

    \newcommand*{\scinot}[2]{#1\times10^{#2}}
    \newcommand*{\rd}[2]{\frac{\mathrm{d}#1}{\mathrm{d}#2}}
    \newcommand*{\rtd}[2]{\frac{\mathrm{d}^2#1}{\mathrm{d}#2^2}}
    \newcommand*{\pd}[2]{\frac{\partial#1}{\partial#2}}
    
    \newcommand*{\mdil}[2]{\mathrm{D}#1/\mathrm{D}#2}
    \newcommand*{\pdil}[2]{\partial#1/\partial#2}
    \newcommand*{\rdil}[2]{\mathrm{d}#1/\mathrm{d}#2}
    \newcommand*{\md}[2]{\frac{\mathrm{D}#1}{\mathrm{D}#2}}
    \newcommand*{\at}[1]{\left.#1\right|}
    \newcommand*{\abs}[1]{\left|#1\right|}
    \newcommand*{\ev}[1]{\left\langle#1\right\rangle}
    \newcommand*{\p}[1]{\left(#1\right)}
    \newcommand*{\s}[1]{\left[#1\right]}
    \newcommand*{\z}[1]{\left\{#1\right\}}
    \newcommand*{\bm}[1]{\mathbf{#1}}
    \newcommand*{\uv}[1]{\hat{\mathbf{#1}}}
    \DeclareMathOperator*{\med}{med}
    \DeclareMathOperator*{\erf}{erf}

    \colorlet{Corr}{red}

\title[Physics of Tidal Dissipation]{Physics of Tidal Dissipation in Early-Type
Stars and White Dwarfs: Hydrodynamical Simulations of Internal Gravity Wave
Breaking in Stellar Envelopes}
\author[Y. Su et\ al.]{
Yubo Su,$^1$,
Daniel Lecoanet,$^{2,3}$
Dong Lai$^{1,4, 5}$
\\
$^1$ Cornell Center for Astrophysics and Planetary Science, Department of
Astronomy, Cornell University, Ithaca, NY 14853, USA
\\
$^2$ Princeton Center for Theoretical Science, Princeton University, Princeton,
NJ 08544, USA
\\
$^3$ Department of Astrophysical Sciences, Princeton University, Princeton NJ
08544, USA
\\
$^4$ Tsung-Dao Lee Institute, Shanghai Jiao Tong University, Shanghai, 200240,
China
\\
$^5$ Department of Astronomy and Miller Institute for Basic Research In Science,
UC Berkeley, Berkeley, CA 94720, USA
}

\date{Accepted 2020 May 06. Received 2020 April 13; in original form 2020
February 26}

\pubyear{2020}

\begin{document}\label{firstpage}
\pagerange{\pageref{firstpage}--\pageref{lastpage}}
\maketitle


\begin{abstract}
    In binaries composed of either early-type stars or white
    dwarfs, the dominant tidal process involves the excitation of internal
    gravity waves (IGWs), which propagate towards the stellar surface, and their
    dissipation via nonlinear wave breaking. We perform 2D hydrodynamical
    simulations of this wave breaking process in a stratified, isothermal
    atmosphere. We find that, after an initial transient phase, the dissipation
    of the IGWs naturally generates a sharp critical layer, separating the lower
    stationary region (with no mean flow) and the upper ``synchronized'' region
    (with the mean flow velocity equal to the horizontal wave phase speed).
    While the critical layer is steepened by absorption of these waves, it is
    simultaneously broadened by Kelvin-Helmholtz instabilities such that, in
    steady state, the critical layer width is determined by the Richardson
    criterion. We study the absorption and reflection of incident waves off the
    critical layer and provide analytical formulae describing its long-term
    evolution. The result of this study is important for characterizing the
    evolution of tidally heated white dwarfs and other binary stars.
\end{abstract}

\begin{keywords}
white dwarfs -- hydrodynamics -- binaries:close -- waves 
\end{keywords}

\setcitestyle{notesep={ }}

\section{Introduction}\label{s:intro}

The physical processes responsible for tidal evolution in close binaries often
involve the excitation and dissipation of internal waves, going beyond the
``weak friction'' of equilibrium tides \citep[see][for a
review]{ogilvie2014tidal}. In particular, internal gravity waves (IGWs), arising
from buoyancy of stratified stellar fluid, play an important role in several
types of binary systems. In solar-type stars with radiative cores and convective
envelopes, IGWs are excited by tidal forcing at the radiative-convective
boundary and propagate inward; as the wave amplitude grows due to geometric
focusing, nonlinear effects can lead to efficient damping of the wave
\citep{goodman1998dynamical,barker_ogilvie,essick2015orbital}. In early-type
main-sequence stars, with convective cores and radiative envelopes, IGWs are
similarly excited at the convective-radiative interface but travel toward the
stellar surface; nonlinearity develops as the wave amplitude grows, leading to
efficient dissipation \citep{zahn75,zahn77}. As the outgoing wave deposits its
angular momentum to the stellar surface layer, a critical layer may form and the
star is expected to synchronize from outside-in \citep{gn89}.

Tidal dissipation can also play an important role in compact double white dwarf
(WD) binary systems (with orbital periods in the range of minutes to hours).
Such binaries may produce a variety of exotic astrophysical systems and
phenomena, ranging from isolated sdB/sdO stars, R CrB stars, AM CVn binaries,
high-mass neutron stars and magnetars (created by the accretion-induced collapse
of merging WDs), and various optical transients (underluminous supernovae,
Ca-rich fast transients, and type Ia supernovae)
\citep[e.g.][]{livio2018progenitors,toloza2019understanding}. The outcomes of WD
mergers depend on the WD masses and composition, but tidal dissipation can
strongly affect the pre-merger conditions of the WDs and therefore the merger
outcomes. Tidal dissipation may also influence the evolution of eccentric
WD-massive black hole binaries prior to the eventual tidal disruption of the WD
\citep{vick2017tidal}.

Recent studies have identified nonlinear dissipation of IGWs as the key tidal
process in compact WD binaries \citep{fullerII,fullerIV,tidal_novae,bukart}:
IGWs are tidally excited mainly at the composition transitions of the WD
envelope; as these waves propagate outwards towards the WD surface, they grow in
amplitude until they break, and transfer both energy and angular momentum from
the binary orbit to the outer envelope of the WD\@. However, these previous
works parameterized the wave breaking process in an ad hoc manner. The details
of dissipation, namely the location and spatial extent of the wave breaking,
affect the observable outcomes: dissipation near the surface of the WD can be
efficiently radiated away and simply brightens the WD, while dissipation deep in
the WD envelope causes an energy buildup that results in energetic flares
\citep{tidal_novae}. An important goal of this paper is to elucidate the details
of the nonlinear IGW breaking process; the result of this ``microphysics'' study
will help determine the thermal evolution and the observational manifestations
of tidally heated binary WDs.

In this paper, we perform numerical simulations of IGW breaking in a
plane-parallel stratified atmosphere (a simple model for a stellar envelope). We
use the pseudo-spectral code Dedalus \citep{dedalus,dedalus2} and a 2D Cartesian
geometry, and consider IGWs propagating into an isothermal fluid initially at
rest. We find that, after an initial transient phase, a \emph{critical layer}
naturally develops, separating a lower zone that has no horizontal mean flow and
an upper zone with mean flow at the horizontal phase velocity of the IGW\@. The
major part of our paper is dedicated to characterizing the behavior of the
critical layer when interacting with a continuous train of IGW excited from the
bottom of the atmosphere. IGWs are generally \emph{anti}-diffusive, in that they
steepen shear flows \citep{lindzen_qbo,lecoanet_meanflow} and act to narrow the
critical layer. We find this steepening is counter-balanced by the
Kelvin-Helmholtz instability and turbulence within the narrow critical layer. By
careful accounting of the momentum flux budget about the critical layer, we are
able to model the reflection and absorption of the incident IGW, and the slow
downward propagation of the critical layer.

While the motivation of our study is to understand tidal dissipation in WD and
early-type stellar binaries, the IGW breaking process studied in this paper is
also quite relevant to the circulation dynamics of planetary atmospheres
\citep[see e.g.][]{lindzen1981turbulence,holton1983influence,baldwin2001quasi}.

This paper is organized as follows. In Section~\ref{s:equations} we present the
system of equations used in our simulations. In Section~\ref{s:theory}, we
review the existing understanding of wave breaking and present analytical
results characterizing IGW behavior near a critical layer. In
Section~\ref{s:numerics} we describe our numerical setup and in
Section~\ref{s:weak_sim} we validate our method in the weak-forcing limit
against linear theory. In Section~\ref{s:sim}, we present the results of
simulations of IGW breaking and our characterization of the critical layer. We
summarize and conclude in Section~\ref{s:discussion}.

\section{Problem Setup and Equations}\label{s:equations}

We consider a incompressible, isothermally stratified fluid representing a
stellar envelope or atmosphere. We study dynamics in 2D, so that fluid variables
depend only on the Cartesian coordinates $x$ and $z$. While it
is well known that waves break differently in 2D versus 3D \citep{klostermeyer,
winters1994}, the dynamical effect of the breaking process is likely to be
similar in 2D \citep{barker_ogilvie}. We approximate the gravitational field as
uniform, pointing in the $(-\uv{z})$ direction. The plane-parallel approximation
is justified since wave breaking generally occurs near the stellar surface. The
background density stratification is given by
\begin{equation}
    \overline{\rho} = \overline{\rho}_0 e^{-z/H},
\end{equation}
with $\overline{\rho}_0$ some reference density. We denote background quantities
with overbars and perturbation quantities with primes.

The Euler equations for an incompressible fluid in a uniform gravitational field
are
\begin{subequations}\label{se:nl_orig}
    \begin{align}
        \bm{\nabla} \cdot \bm{u} &= 0,\label{eq:nl_incomp}\\
        \md{\rho}{t} &= 0 ,\label{eq:nl_density}\\
        \md{\bm{u}}{t} + \frac{\bm{\nabla}P}{\rho} + g\uv{z} &=
            0\label{eq:nl_mom},
    \end{align}
\end{subequations}
where $\mdil{}{t} = \pdil{}{t}\,+\,\p{\bm{u} \cdot \bm{\nabla}}$ is the
Lagrangian or material derivative, and $\bm{u}, \rho, P$ denote the velocity
field, density and pressure respectively. The constant gravitational
acceleration is $(-g\uv{z})$. Note that these equations
conserve the same wave energy as the commonly used anelastic equations
\citep{ogura1962scale,anel_part1} and thus give the same wave amplitude
growth. Appendix~\ref{s:equation_deriv} provides a derivation of these
equations and justification for using them.

For this isothermal background, hydrostatic equilibrium implies
$\overline{P}(z) = \overline{\rho}(z) g H$. We assume there is
initially no background flow, so $\bm{u} = \bm{u}'$. Physically, this assumption
corresponds to a non-rotating star.

For convenience, we introduce the dimensionless density variable $\Upsilon$ and
the reduced pressure $\varpi$ \citep[e.g.][]{lecoanet_anel} via
\begin{align}
    \Upsilon &\equiv \ln \frac{\rho}{\bar{\rho}},\\
    \varpi &\equiv \frac{P}{\rho}.
\end{align}
These variables automatically enforce $\rho > 0$ and eliminate the stiff term
$\bm{\nabla} P / \rho$ in the Euler equation. In terms of $\Upsilon$ and
$\varpi$, the second two equations in~\eqref{se:nl_orig} become
\begin{subequations}\label{se:nl_upsilon}
    \begin{align}
        \md{\Upsilon}{t} + u_z \pd{\ln \overline{\rho}}{z} &= 0
            ,\label{eq:nl_up_density} \\
        \md{\bm{u}}{t} + \bm{\nabla}\varpi + \varpi\bm{\nabla}\Upsilon
            - \frac{\varpi}{H}\uv{z} + g\uv{z} &= 0\label{eq:nl_upsilon_u}.
    \end{align}
\end{subequations}
Hydrostatic equilibrium corresponds to $\Upsilon = 0, \overline{\varpi} = gH$.

\section{Internal Gravity Waves: Theory}\label{s:theory}

\subsection{Linear Analysis}\label{ss:lin_analysis}

In the small perturbation limit, we may linearize Eq.~\eqref{se:nl_upsilon}.
The resulting equations admit the canonical IGW solution
\citep{drazin,sutherland0}
\begin{equation}
    u_z'\p{x, z, t} = Ae^{z/2H}\cos\p{k_{x}x + k_{z}z - \omega t},
        \label{eq:lin_sol}
\end{equation}
where $A$ is a constant amplitude, and the frequency $\omega$ and the wave
number $\p{k_x, k_z}$ satisfy the dispersion relation
\begin{equation}
    \omega^2 = \frac{N^2k_{x}^2}{k_{x}^2 + k_{z}^2 + \p{2H}^{-2}}.
        \label{eq:disp_rel}
\end{equation}
Our equations are valid in the limit of large sound speed ($c_s \to \infty$), in
which the \emph{Brunt-V\"ais\"al\"a frequency}, $N$, is given by
\begin{equation}
    N^2 \equiv g^2\p{\rd{\rho}{P} - \frac{1}{c_s^2}} = \frac{g}{H},
\end{equation}
and is constant. Other dynamical quantities are simply related to $u'_z$.

In the short-wavelength/WKB limit ($\abs{k_{z}H} \gg 1$), the solution exhibits
the following characteristics:
\begin{enumerate}
    \item The amplitude of the wave grows with $z$ as $e^{z/2H}$. Thus, the
        linear approximation always breaks down for sufficiently large $z$.

    \item The phase and group velocities are given by:
        \begin{align}
            \bm{c}_{p} &=
                \p{k_{x}\uv{x} + k_{z}\uv{z}}\frac{\omega}
                {k_{x}^2 + k_{z}^2 + \p{2H}^{-2}},\\
            \bm{c}_{g} &= N\frac{\s{k_{z}^2 + \p{2H}^{-2}}\uv{x}
                - \p{k_{x}k_{z}\uv{z}}}
                {\s{k_{x}^2 + k_{z}^2 + \p{2H}^{-2}}^{3/2}}.\label{eq:vg}
        \end{align}
        The additional $\p{2H}^{-2}$ term in the denominator
        accounts for the growing amplitude of the IGW in the $z$ direction (as
        the $z$ wavenumber is effectively $k_z - i / (2H)$). We note
        $\bm{c}_{p} \cdot \bm{c}_g = \mathcal{O}\s{\p{k_{z}H}^{-2}} \approx 0$.
        In the Boussinesq approximation where terms of order
        $\mathcal{O}\p{H^{-2}}$ are ignored, the phase and group velocities are
        exactly orthogonal \citep{drazin,sutherland1}. We use the convention
        where upward propagating IGW have $c_{g, z} > 0$, $k_z < 0, k_x > 0$.

    \item The averaged horizontal momentum flux $F$ (in the $+\uv{z}$
        direction) carried by the IGW is defined by
        \begin{equation}
            F(z, t) \equiv \ev{\rho u_{x}' u_{z}'}_x \equiv
                \frac{1}{L_x}\int_0^{L_x}\limits \rho u_{x}'u_{z}'\;\mathrm{d}x.
                    \label{eq:F_def}
        \end{equation}
        The notation $\ev{\dots}_x$ denotes averaging over the $x$ direction.
        For the linear solution (Eq.~\ref{eq:lin_sol}), this evaluates to
        \begin{equation}
            F \approx -\frac{A^2}{2}\overline{\rho}_0\frac{k_{z}}{k_{x}},
                    \label{eq:S_lin}
        \end{equation}
        Thus, indeed $F > 0$ for an upward propagating IGW ($c_{g, z} > 0$).
\end{enumerate}

\subsection{Wave Generation}\label{ss:forcing}

To model continuous excitation of IGWs deep in the stellar envelope propagating
towards the surface, we use a volumetric forcing term to excite IGW near the
bottom of the simulation domain. Our forcing excites both IGWs
propagating upwards, imitating a wave tidally excited deeper in the star, and
downwards, which are not physically relevant in binaries. In our simulations,
these downward propagating waves are dissipated by a damping zone described in
Section~\ref{ss:damping}.

As not to interfere with the incompressibility constraint, we force the system
on the density equation. We implement forcing with strength $C$ localized around
height $z_0$ with small width $\sigma$ by replacing Eq.~\eqref{eq:nl_up_density}
with
\begin{equation}
    \md{\Upsilon}{t} + u_{z}\pd{\ln \overline{\rho}}{z}
        = Ce^{-\frac{(z - z_0)^2}{2\sigma^2}}
            \cos \p{k_{x}x - \omega t}.\label{eq:vol_drive}
\end{equation}
Using a narrow Gaussian profile excites a broad $z$ power spectrum, but only the
$k_{z}$ satisfying the dispersion relation (Eq.~\ref{eq:disp_rel}) for the given
$k_{x}$ and $\omega$ will propagate.

In the linearized system, the effect of this forcing can be solved exactly
(see Appendix~\ref{s:force_solved}). In the limit $\abs{k_zH} \gg 1, \sigma
\ll H$, the solution can be approximated as two plane waves propagating away
from the forcing zone
\begin{align}
    u_{z}&(x, z, t) \approx{} \frac{C}{2k_z}\frac{gk_x^2}{\omega^2}
        \exp\p{-\frac{k_z^2\sigma^2}{2}}
        \sqrt{2\pi \sigma^2} \nonumber\\
        &{}\times\begin{cases}
        e^{\frac{z - z_0}{2H}}\sin\p{k_{x}x + k_{z}(z - z_0) - \omega t
            + \frac{k_z\sigma^2}{2H}}
            & \text{for }z > z_0,\\[5pt]
        e^{\frac{z - z_0}{2H}}\sin\p{k_{x}x - k_{z}(z - z_0) - \omega t
            + \frac{k_z\sigma^2}{2H}}
            & \text{for }z < z_0.\\
    \end{cases}\label{eq:uz_lin}
\end{align}
The $z > z_0$ region contains an upward propagating IGW wavetrain. The $x$
component of the velocity can be obtained by the incompressibility constraint
(Eq.~\ref{eq:nl_incomp}).

\subsection{Wave Breaking Height}\label{ss:wave_breaking}

As the upward propagating IGW grows in amplitude ($\abs{\bm{u}} \propto
e^{z/2H}$), it is expected to break due to nonlinear effects. We can estimate
the height of wave breaking using the condition $\abs{\bm{u}} \sim \omega /
\abs{\bm{k}}$. This can be rewritten using the Lagrangian displacement
$\boldsymbol{\xi} = \bm{u} / \p{-i\omega}$:
\begin{equation}
    \abs{\xi_z k_z} \sim 1.\label{eq:nl}
\end{equation}

\citet{drazin, klostermeyer, winters1994} describe the onset of wave breaking in
some detail. At intermediate amplitudes, wave breaking occurs via triadic
resonances, transferring energy from the ``parent'' IGW to ``daughter'' waves on
smaller length scales that efficiently damp. The horizontal momentum flux
decreases from $F$ to $0$ over this breaking region. The lost flux is deposited
into a horizontal mean flow
\begin{equation}
    \overline{U}(z, t) \equiv \ev{u_x}_x.\label{eq:mean_flow}
\end{equation}
As the mean flow grows, a \emph{critical layer} may form, as discussed below.

\subsection{Critical Layers}\label{ss:crit_layer}

A horizontal shear flow $\overline{U}(z, t)\uv{x}$ enters the fluid equations
via the Lagrangian derivative, which can be decomposed as
\begin{equation}
    \md{}{t} = \pd{}{t} + \overline{U} \pd{}{x} + \p{\bm{u}' \cdot \bm{\nabla}},
\end{equation}
where $\bm{u}'$ is the velocity field \emph{without} the shear flow. Thus,
$\overline{U}$ has the effect of Doppler shifting the time derivative into the
frame comoving with the mean flow. If $\overline{U}$ is roughly constant, then
the behavior of a linear plane-wave perturbation satisfies the modified
dispersion relation
\begin{equation}
    \p{\omega - \overline{U}k_x}^2 =
        \frac{N^2k_{x}^2}{k_{x}^2 + k_{z}^2 + \p{2H}^{-2}}.
        \label{eq:disp_rel_U}
\end{equation}
This is just Eq.~\eqref{eq:disp_rel} with $\omega \to \omega - \overline{U}
k_x$. It is apparent that if $\overline{U} = \overline{U}_c$, where
\begin{equation}
    \overline{U}_c \equiv \frac{\omega}{k_x},\label{eq:u_crit}
\end{equation}
then the dispersion relation is singular and the linear solution breaks down.
Physically, this corresponds to the Doppler-shifted frequency of the IGW being
zero. Anywhere $\overline{U} = \overline{U}_c$ is called a \emph{critical
layer}.

The behavior of an IGW incident upon a critical layer was first studied in the
inviscid, linear regime in \citet{booker_bretherton}, which found nearly
complete absorption of the IGW\@. The amplitude
reflection and transmission coefficients are given by
\begin{align}
    \mathcal{R} &= \exp\p{-2\pi \sqrt{\mathrm{Ri} - \frac{1}{4}}}, &
    \mathcal{T} &= \exp\p{-\pi \sqrt{\mathrm{Ri} - \frac{1}{4}}},
        \label{eq:crit_coeffs}
\end{align}
where $\mathrm{Ri}$ is the local Richardson number evaluated at the critical
layer height $z_c$:
\begin{equation}
    \mathrm{Ri} \equiv \at{\frac{N^2}{\p{\pdil{\overline{U}}{z}}^2}}_{z_c}.
        \label{eq:ri_def}
\end{equation}
In the $\mathrm{Ri} \gg 1$ limit, $\mathcal{R}, \mathcal{T} \ll 1$ and the
incident wave is almost completely absorbed. This result also applies to viscous
fluids \citep{hazel}. However, weakly nonlinear theory \citep{brown_stewartson}
and numerical simulations \citep{winters1994} suggest that nonlinear effects may
significantly enhance reflection and transmission.

Consider now the long-term evolution of the critical layer due to continuous
horizontal momentum transfer by IGWs. Any incident horizontal
momentum flux absorbed by the fluid, denoted $F_a(t)$, must manifest as
additional horizontal momentum of the shear flow.
Additionally, as the mean flow $\overline{U}$ cannot grow
efficiently above $\overline{U}_c$ (due to the breakdown of the linear
solution), we assume $\overline{U}$ saturates at $\overline{U}_c$, which holds
to good accuracy (see Fig.~\ref{fig:nl_fluxes}). In this case, the critical
layer must propagate downward in response to the incident momentum flux. The
horizontal momentum of the shear flow satisfies
\begin{equation}
    \pd{}{t}\int\limits \overline{\rho}(z) \overline{U}(z, t)\;\mathrm{d}z
        - F_a(t) = 0.
\end{equation}
Assuming $\overline{U}(z > z_c) \approx \overline{U}_c$ and $\overline{U}(z <
z_c) \approx 0$, this condition becomes
\begin{equation}
    -\overline{\rho}(z_c) \overline{U}_c\rd{z_c}{t} = F_a(t).\label{eq:zc_anal}
\end{equation}
If $F_a$ is constant in time, the height of the critical layer $z_c(t)$ has
analytical solution:
\begin{equation}
    z_c(t) = -H\ln \s{\exp\p{-\frac{z_c(t = 0)}{H}} +
        \frac{tF_a}{\overline{U}_c H\overline{\rho}_0}},\label{eq:zc_sol}
\end{equation}
where $z_c(t = 0)$ is the initial critical layer height.

\section{Numerical Simulation Setup}\label{s:numerics}

We use the pseudo-spectral code Dedalus \citep{dedalus,dedalus2} to simulate
the excitation and propagation of IGWs (Section~\ref{s:weak_sim}) as well as
their nonlinear breaking and the formation of a critical layer
(Section~\ref{s:sim}).

\subsection{Parameter Choices}\label{ss:params}

We solve Eqs.~\eqref{eq:nl_incomp},~\eqref{eq:nl_upsilon_u},
and~\eqref{eq:vol_drive} in a Cartesian box with size $L_x, L_z$. We choose
periodic boundary conditions in both the $x$ and $z$ direction. To mimic the
absence of physical boundaries at the top/bottom of the simulation domain, we
damp perturbations to zero near the top/bottom using damping zones (see
Section~\ref{ss:damping}). We expand all variables as Fourier series with $N_x$
and $N_z$ modes, and use the $3/2$ dealiasing rule to avoid aliasing errors in
the nonlinear terms \citep{boyd}. We choose $L_z = 12.5H$ ($z$ runs from $-H$ to
$11.5H$), and the lower and upper damping zones are active for $z < 0.3H$ and $z
> 9.5H$ respectively. The forcing (see Eq.~\eqref{eq:vol_drive}) is centered at
$z_0 = 2H$ with width $\sigma = 0.078H$, sufficiently far from the lower damping
zone and permitting sufficient room for the upward propagating wave to grow as
$\propto e^{z/2H}$. Finally, we want similar grid spacing in the $x$ and $z$
directions (i.e.\ $L_x / N_x \sim L_z /N_z$), guided by the intuition that
turbulence generated by wave breaking is approximately isotropic, so we use $L_x
= 4H$ and $N_z / N_x = 4$.

The time integration uses a split implicit-explicit third-order scheme where
certain terms are treated implicitly and the remaining terms are treated
explicitly. A third-order, four-stage DIRK-ERK scheme \citep{ascher} is used
with adaptive timesteps computed from the minimum of $0.1 / N$
and the advective Courant-Friedrichs-Lewy (CFL) time. The CFL time is given by
$\Delta t = 0.7 \min(\Delta x / u_x,\Delta z / u_{z})$, where the minimum is
taken over every grid point in the domain, and $\Delta x \equiv L_x / N_x$ and
$\Delta z \equiv L_z / N_z$ are the grid spacings in the $x$ and $z$ directions
respectively.

We non-dimensionalize the problem such that $H = N = \rho_0 = 1$. The physics of
the simulation is then fixed by the four remaining parameters $k_{x}$, $\omega$,
$C$, and the viscosity $\nu$. We describe our choices for these parameters
below:
\begin{enumerate}
    \item $k_{x}$: Tidally excited waves in stars generally have $\ell = 2$,
        corresponding to a horizontal wavenumber $k_\perp\sim 1/R$, where $R$ is
        the radius of the star. We use the smallest wavenumber in our
        simulation, $k_x=2\pi/L_x$.

    \item $\omega$: We choose $\omega$ by evaluating the dispersion relation
        $\omega(k_x, k_z)$ for a desired $k_z$ (see Eq.~\eqref{eq:disp_rel}). We
        pick $\abs{k_z H} = 2\pi$ to ensure the waves are very well resolved in
        all of our simulations. Note however that tidally forced IGWs typically
        have $\omega \ll N$, or equivalently $k_r/k_\perp \sim k_r R \gg 1$.
        This requires $\abs{k_z H} \gtrsim 1$, which is only marginally
        satisfied in our simulations.

    \item $C$: In our weak forcing simulations (Section~\ref{s:weak_sim}),
        we first choose the forcing strength $C$ (see Eq.~\eqref{eq:vol_drive})
        to be sufficiently weak such that $\abs{\xi_z k_z} \ll 1$ is satisfied
        everywhere in the simulation domain. This constrains $C$ by
        Eq.~\eqref{eq:uz_lin}. In our wave breaking simulations
        (Section~\ref{s:sim}), we choose larger $C$.

    \item $\nu$: Nonlinear effects transfer wave energy from the injection
        wavenumber $\bm{k}$ to larger wavenumbers. Our spectral method does not
        have any numerical viscosity, so diffusivity must be introduced into the
        equations to regularize the systems at large wavenumbers. We add
        viscosity and diffusivity to the system in a way that conserves
        horizontal momentum (see Appendix~\ref{se:strat_impl} for details). We
        define the dimensionless Reynolds number
        \begin{equation}
            \mathrm{Re} \equiv \frac{\omega}{\nu k_{z}^2}
                = \frac{\omega}{\nu}\p{\frac{H}{2\pi}}^2.\label{eq:re_def}
        \end{equation}
        We use $\mathrm{Re} \gg 1$ in our simulations\footnote{This condition
        is always satisfied in stars. For example, in WDs, the dominant linear
        dissipation mechanism of g-modes is radiative damping, with damping rate
        ranging from $10^{-11}$--$10^{-4}$ of the mode frequency
        \citep{fullerI}. This corresponds to a small effective viscosity or
        $\mathrm{Re} \gg 1$.}.
\end{enumerate}

Finally, we use initial conditions $\bm{u}(x, z, 0) = \Upsilon(x, z, 0) = 0$ and
$\varpi(x, z, 0) = 1$, corresponding to hydrostatic equilibrium and no initial
fluid motion.

\subsection{Damping Layers}\label{ss:damping}

We aim to damp disturbances that reach the vertical boundaries of the simulation
domain without inducing nonphysical reflection. To do so, we replace material
derivatives in Eq.~\eqref{se:nl_upsilon} with:
\begin{align}
    \md{}{t} &\to \md{}{t} + \Gamma(z),\\
    \Gamma(z) &= \frac{1}{2\tau}\s{2 + \tanh \frac{z - z_T}{\Delta z}
        + \tanh \frac{z_B - z}{\Delta z}},\label{eq:Gamma}
\end{align}
where $z_B = 0.3H$ and $z_T = 9.5H$ are the boundaries of the lower and upper
damping zones respectively. This damps perturbations below $z_B$ and
above $z_T$ with damping time $\tau$ and negligibly affects the dynamics between
$z_B$ and $z_T$. Most importantly, horizontal momentum remains
conserved between $z_B$ and $z_T$, and outgoing boundary conditions are imposed
at $z_B, z_T$. We choose the transition width $\Delta z = 0.25H$ and damping
time $\tau = 1 / (15N)$. This prescription is similar to \citet{lecoanet_damp}
and has the advantage of being smooth, important for spectral methods. Further
details of our implementation of the fluid equations in Dedalus are described in
Appendix~\ref{se:strat_impl}.

\section{Weakly Forced Numerical Simulation}\label{s:weak_sim}

To test our numerical code and implementation, we carry out a simulation in the
weakly forced regime with $C = \scinot{1.64}{-7}$. According to the linear
solution (Eq.~\eqref{eq:uz_lin}), this generates IGW with $\abs{\xi_z k_z}
\approx \scinot{5}{-5}$ just above the forcing zone. The IGW grows to $\abs{\xi_z
k_z} \approx \scinot{7.4}{-3}$ at the upper damping zone and satisfies
$\abs{\xi_z k_z} \ll 1$ in the entire simulation domain. We include a nonzero
$\nu$ corresponding to $\mathrm{Re} = 10^7$.

We expect the waves to follow the analytical solution given by
Eq.~\eqref{eq:uz_lin} and the corresponding $u_x(x, z, t)$; we denote this
analytical solution $\bm{u}_{an}(x, z, t)$. The amplitude of the observed IGW in the
simulation field $\bm{u}$ relative to $\bm{u}_{an}$ over
some region $z \in [z_b, z_t]$ can be estimated from
\begin{equation}
    A_i(t) = \frac{\int\limits_{z_b}^{z_t}\int\limits_0^{L_x}
        \overline{\rho}\p{\bm{u} \cdot \bm{u}_{an}}\;\mathrm{d}x\mathrm{d}z}
        {\int\limits_{z_b}^{z_t}\int\limits_0^{L_x}
        \overline{\rho}\abs{\bm{u}_{an}}^2\;\mathrm{d}x\mathrm{d}z}.
        \label{eq:ahat_def}
\end{equation}
The subscript $i$ denotes the incident wave. If $\bm{u} = \bm{u}_{an}$, then
$A_i(t) = 1$. The factor of $\bar{\rho}$ inside the integrands
in Eq.~\eqref{eq:ahat_def} corrects for the $\propto e^{z/2H}$ growth of
$\bm{u}_{an}$; without it, $A_i(t)$ would be dominated by the contribution near
$z_t$.

\begin{figure}
    \centering
    \includegraphics[width=0.9\columnwidth]{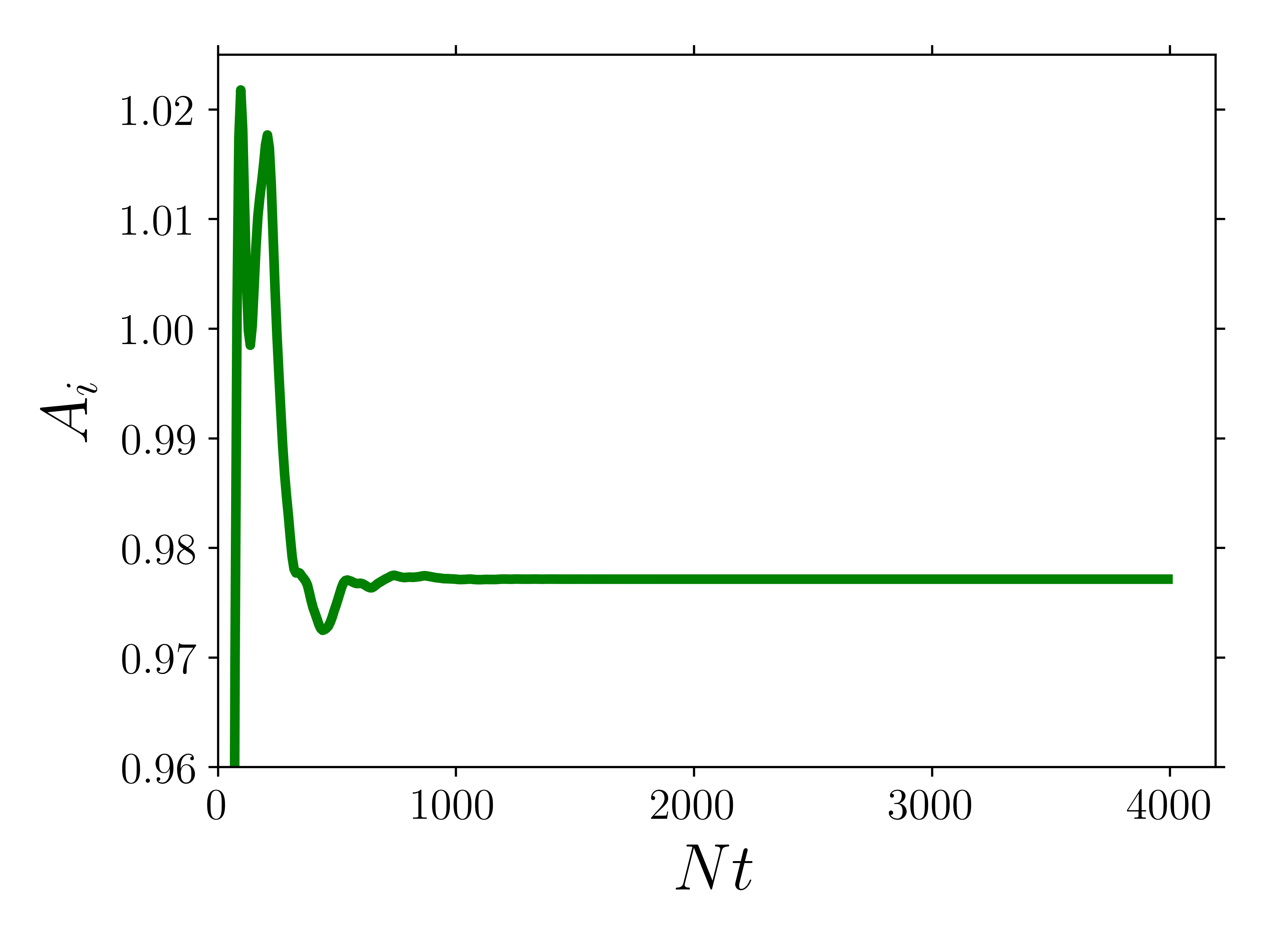}
    \caption{Amplitude of the excited IGW over time (in units of $N^{-1}$) in
    the weakly forced simulation, computed using Eq.~\eqref{eq:ahat_def}.
    $A_i(t) = 1$ corresponds to perfect agreement with the analytical estimate.
    After an initial transient phase, we observe $A_i(t)$ asymptotes to $\approx
    1$, implying continuous excitation of identical IGW with the expected
    amplitude. The small deviation of $A_i(t)$ from unity may be due to
    truncation error in our implicit timestepping scheme, as a
    relatively large fixed step size $\Delta t = 0.1/N$ was used for this
    simulation.}\label{fig:lin_amps}
\end{figure}

For the weakly forced simulation, we expect
$A_i(t) = 1$ when integrated between the forcing and damping zones, i.e.\
$z_b \gtrsim z_0$ and $z_t \lesssim z_T$ ($z_0, z_T$ are defined in
Eq.~\eqref{eq:vol_drive} and Eq.~\eqref{eq:Gamma} respectively). For consistency
with the nonlinear case later, we choose $z_b = z_0 + 3\sigma$ and $z_t = z_b +
H$. Note that using a larger integration domain by choosing $z_t = z_T - \Delta
z$ just below the upper damping zone instead does not change the measured $A_i$.
The resulting measurement of $A_i(t)$ is shown in Fig.~\ref{fig:lin_amps}, and
indeed $A_i \approx 1$ after the initial transient.

\begin{figure}
    \centering
    \includegraphics[width=0.9\columnwidth]{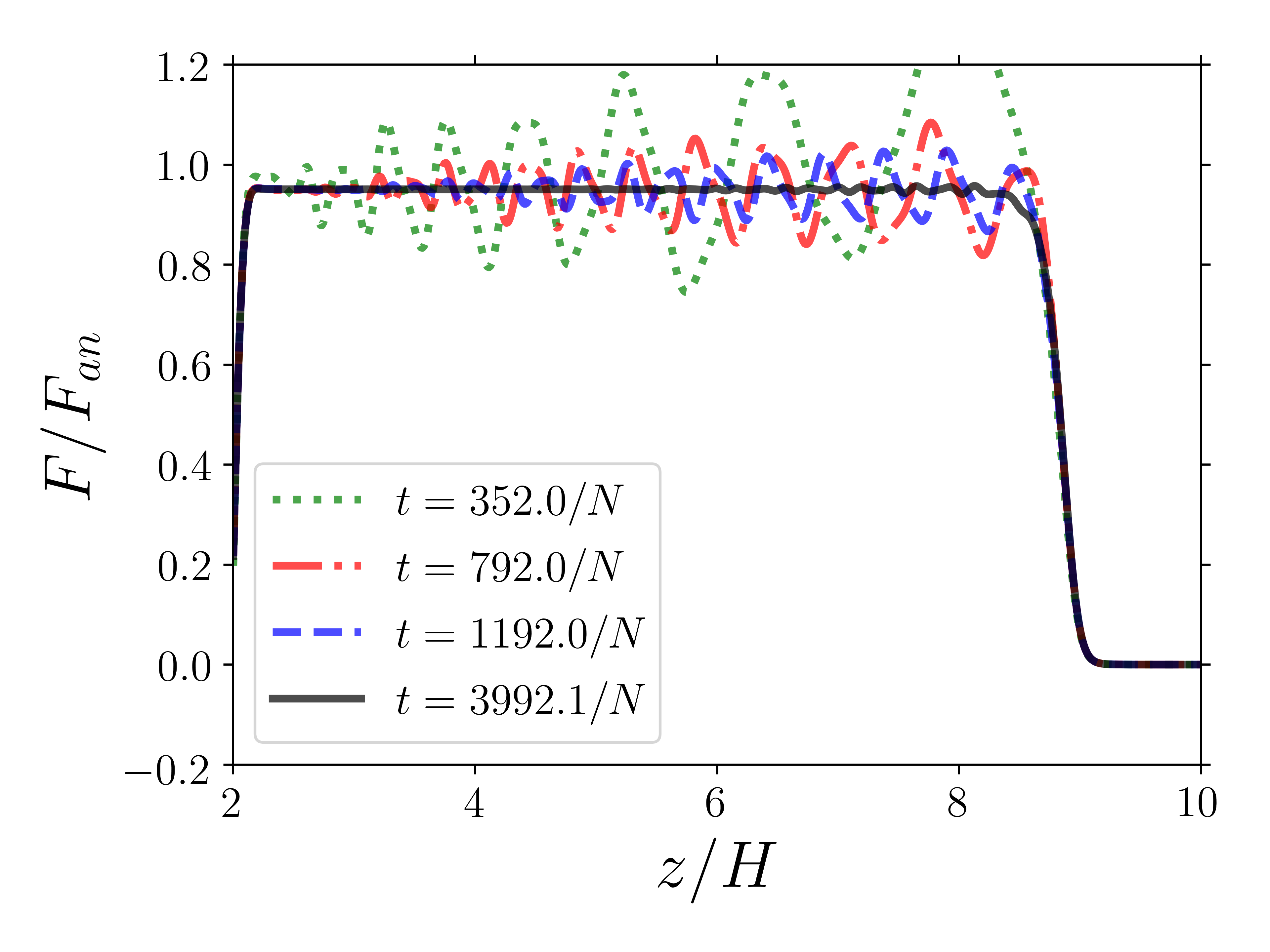}
    \caption{$F/F_{an}$ plotted at select times $t$ (in units of
    $N^{-1}$). As the initial transient dies out, $F / F_{an} \approx 1$ to a
    good approximation above the forcing zone $z > z_0 = 2H$ and below the
    damping zone $z \lesssim z_T = 9.5H$. The horizontal momentum flux excited
    in the forcing zone is transported without loss to the top of the domain,
    where it is dissipated by the damping zone (see Section~\ref{ss:damping})
    without reflection.}\label{fig:lin_fluxes}
\end{figure}

The analytical theory (Section~\ref{ss:lin_analysis}) also predicts that the
horizontal momentum flux $F(z, t)$ is independent of $z$ between the forcing
zone where the wave is generated and the damping zone where it is dissipated.
The expected horizontal momentum flux carried by the excited IGW in the linear
theory can be computed by simply evaluating Eq.~\eqref{eq:F_def} for
$\bm{u}_{an}$ and is a constant:
\begin{equation}
    F_{an} \equiv \ev{\rho u_{an, x} u_{an, z}}_x.\label{eq:F_al}
\end{equation}
Denoting the momentum flux measured in the simulation by $F(z, t)$
(use Eq.~\eqref{eq:F_def} with velocities taken from the simulation), we expect
$F(z, t) = F_{an}$ between $z_0$ and $z_T$. Fig.~\ref{fig:lin_fluxes} shows
agreement with this prediction.

\section{Numerical Simulations of Wave Breaking}\label{s:sim}


To perform simulations of wave breaking phenomena, we use the same setup as
described in Section~\ref{s:numerics} and Section~\ref{s:weak_sim} except for
different values of $C$ and $\nu$. In particular, we choose $C$ such that
$\abs{\xi_z k_z} = 0.1$ in the forcing zone ($z = z_0$). The linear solution
predicts $\abs{\xi_z k_z} \sim 4.25$ at the upper damping zone $z_T$.
We choose the viscosity $\nu$ for each resolution to be as
small as possible while still resolving the shortest spatial scales of the wave
breaking. A table of our simulations can be found in Table~\ref{tab:params}.
\begin{table}
    \centering
    \begin{tabular}{l c c c}
        Resolution & $\mathrm{Re}$\\\bottomrule
        $1024 \times 4096$ & $2048$\\
        $768 \times 3072$ & $1024$\\
        $512 \times 2048$ & $512$\\
        $256 \times 1024$ & $341$\\
        $256 \times 1024$ & $205$\\
        $256 \times 1024$ & $146$\\
    \end{tabular}
    \caption{Spectral resolutions and Reynolds numbers of simulations of wave
    breaking.}\label{tab:params}
\end{table}

\subsection{Numerical Simulation Results}\label{ss:nl_ns}

A full video of our simulation with $N_x = 768$, $N_z = 3072$, $\mathrm{Re} =
1024$ is available
online\footnote{\url{https://academic.oup.com/mnras/article-abstract/495/1/1239/5835700\#supplementary-data}}.
We take this to be our fiducial simulation for the remainder of this paper,
though other simulations show qualitatively similar behavior.

\begin{figure*}
    \includegraphics[width=0.45\textwidth]{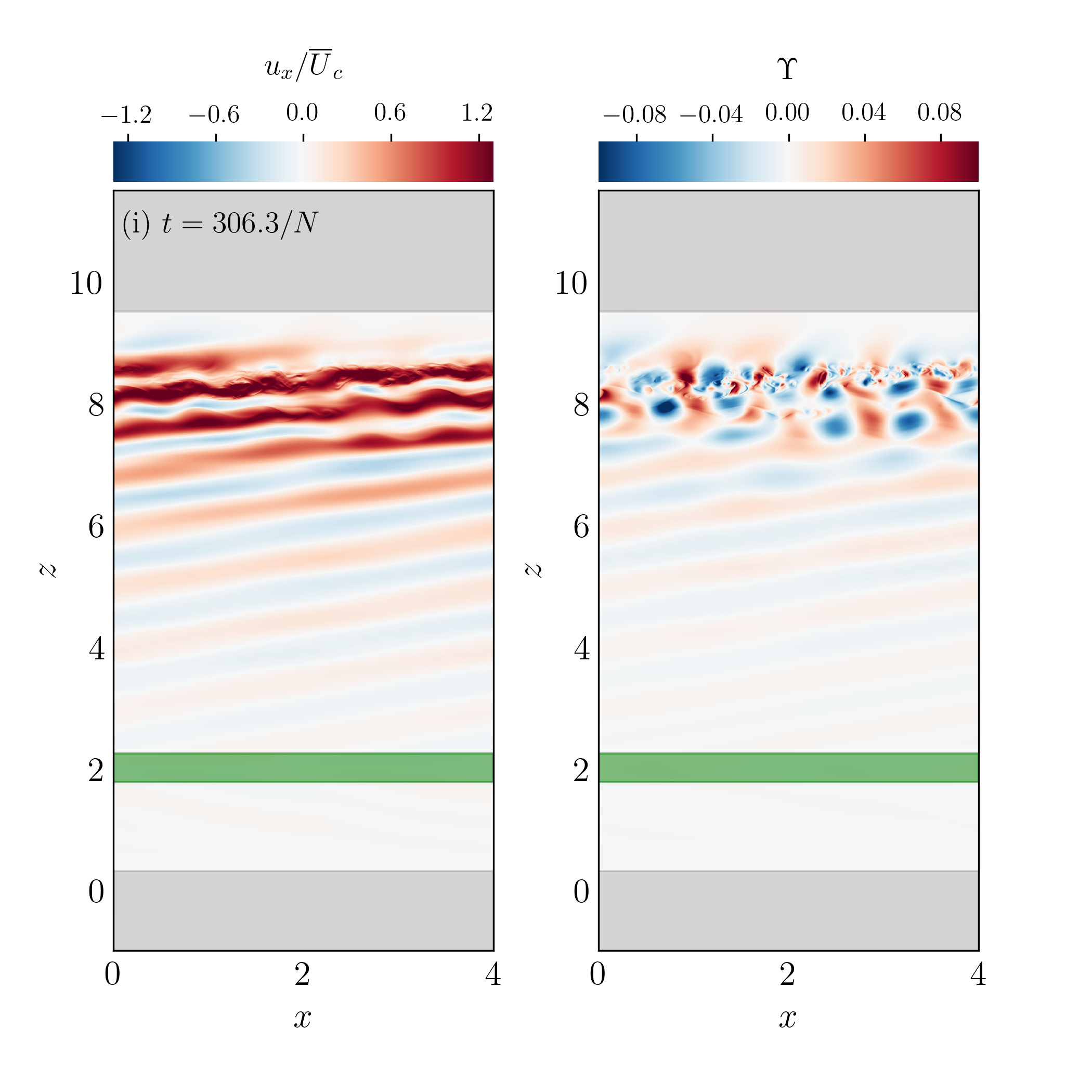}\hfil
    \includegraphics[width=0.45\textwidth]{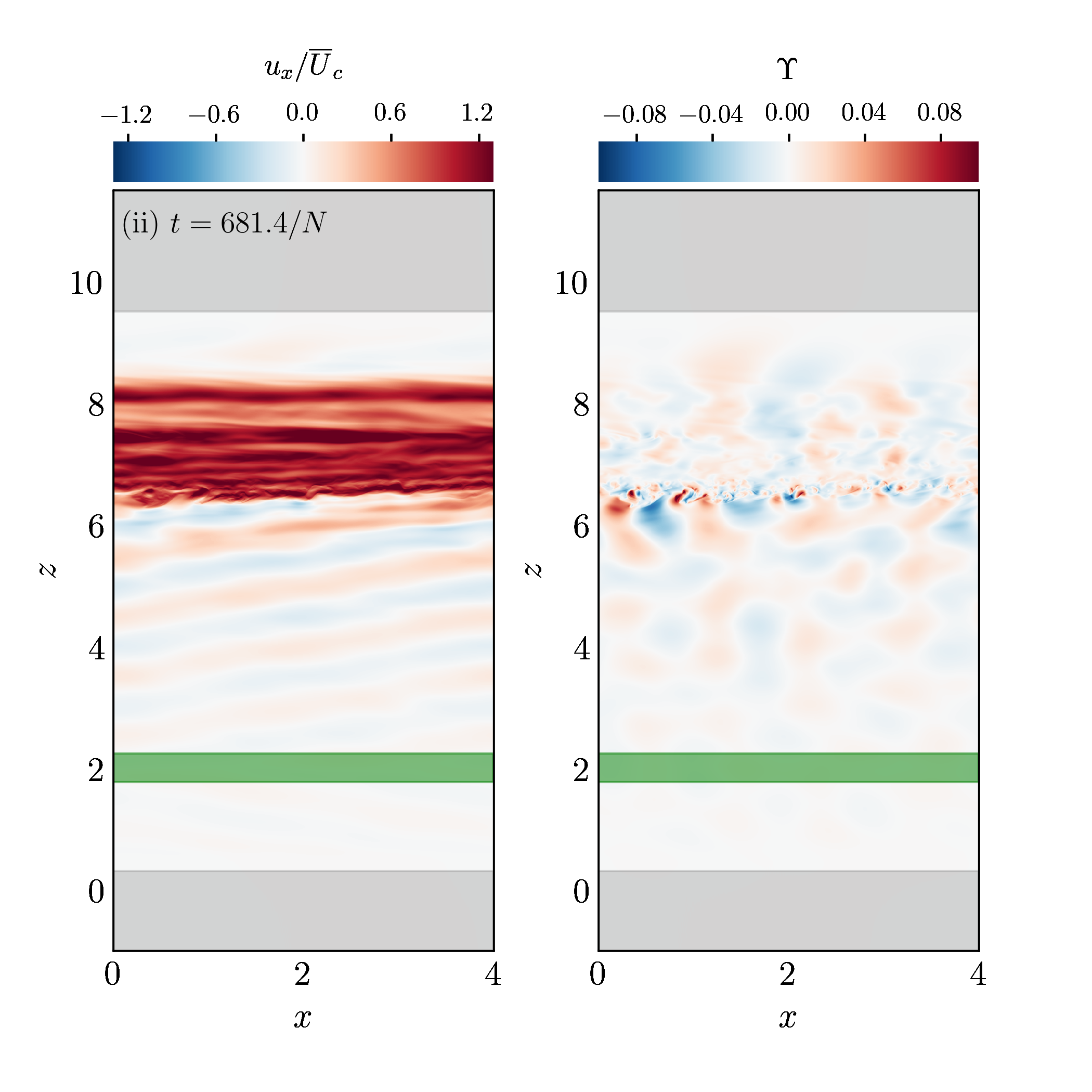}

    \includegraphics[width=0.45\textwidth]{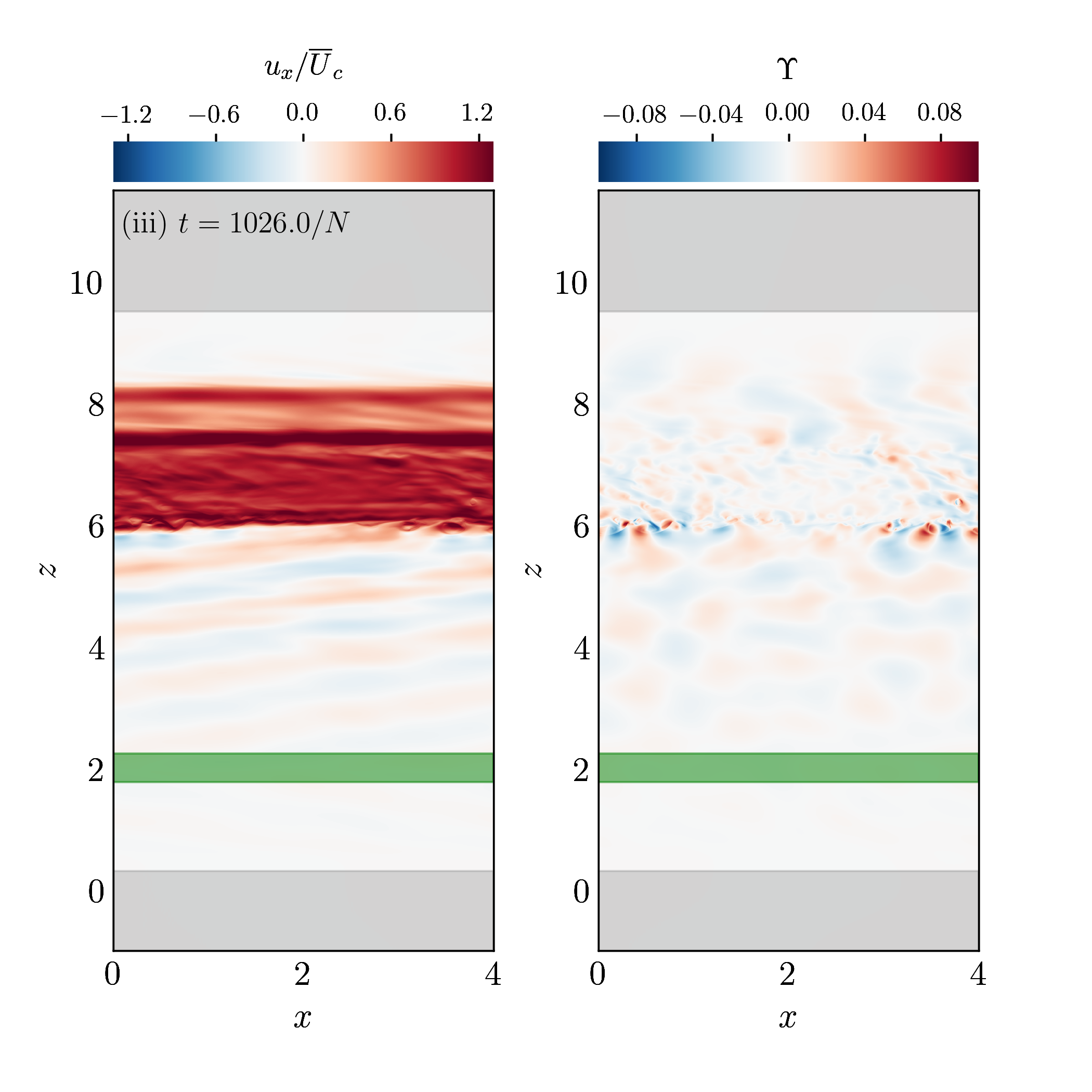}\hfil
    \includegraphics[width=0.45\textwidth]{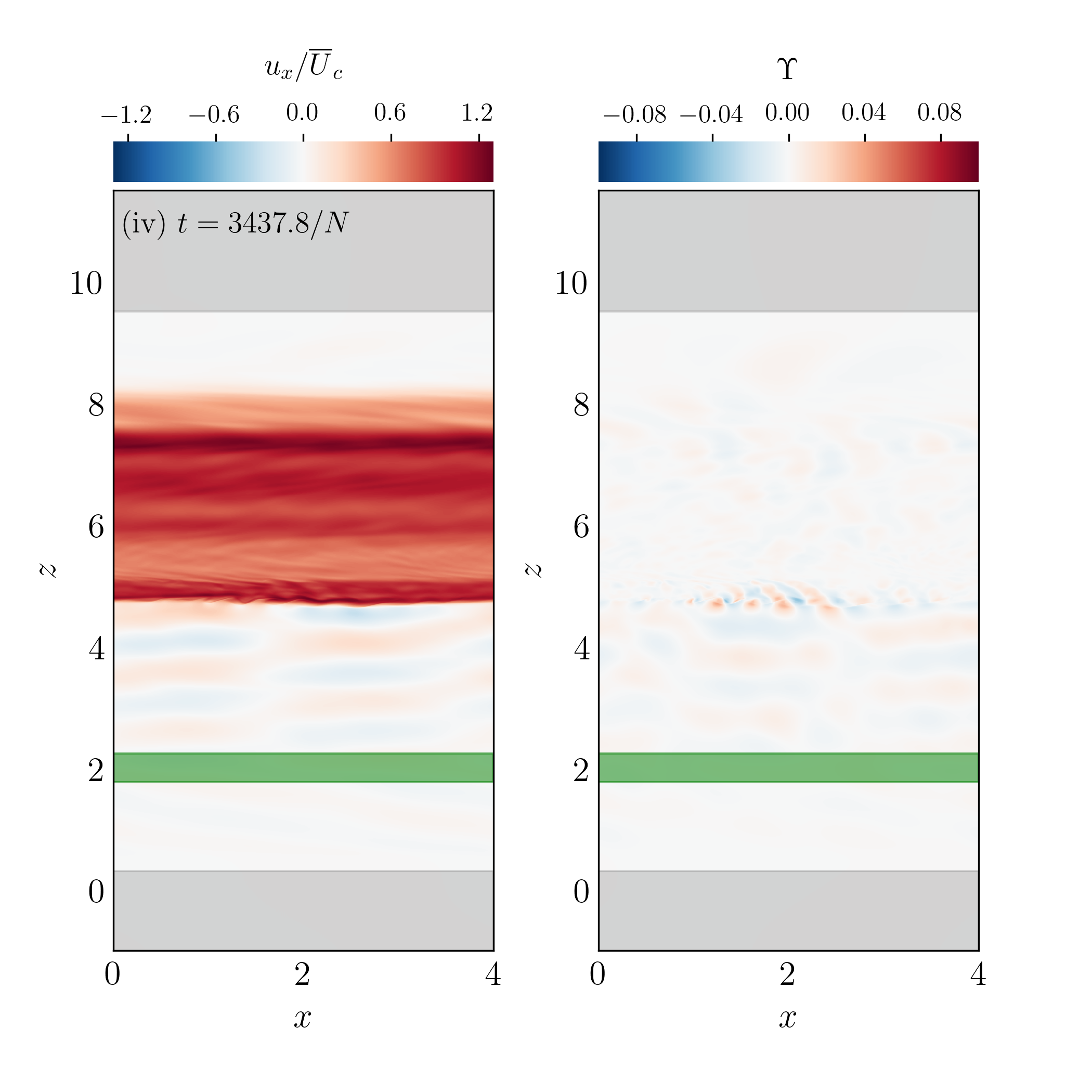}
    \caption{Snapshots of $u_x$ and $\Upsilon \equiv \ln \p{\rho / \bar{\rho}}$
    in the fiducial simulation illustrating distinct phases of the evolution of
    the flow. See the online PDF for a color version. Damping
    layers at the top and bottom of the simulation domain are
    shaded in light grey (see Section~\ref{ss:damping}), while the forcing zone
    in the lower middle portion of the simulation domain (see
    Section~\ref{ss:forcing}) is shaded in light green (boundaries are at $z_0
    \pm 3\sigma$). The four panels illustrate (i) the initial transient wave
    breaking phase, (ii) formation of a distinct critical layer, (iii)
    steepening of the critical layer, and (iv) downward advance of the critical
    layer.}\label{fig:snapshots}
\end{figure*}

\begin{figure}
    \centering
    \includegraphics[width=0.9\columnwidth]{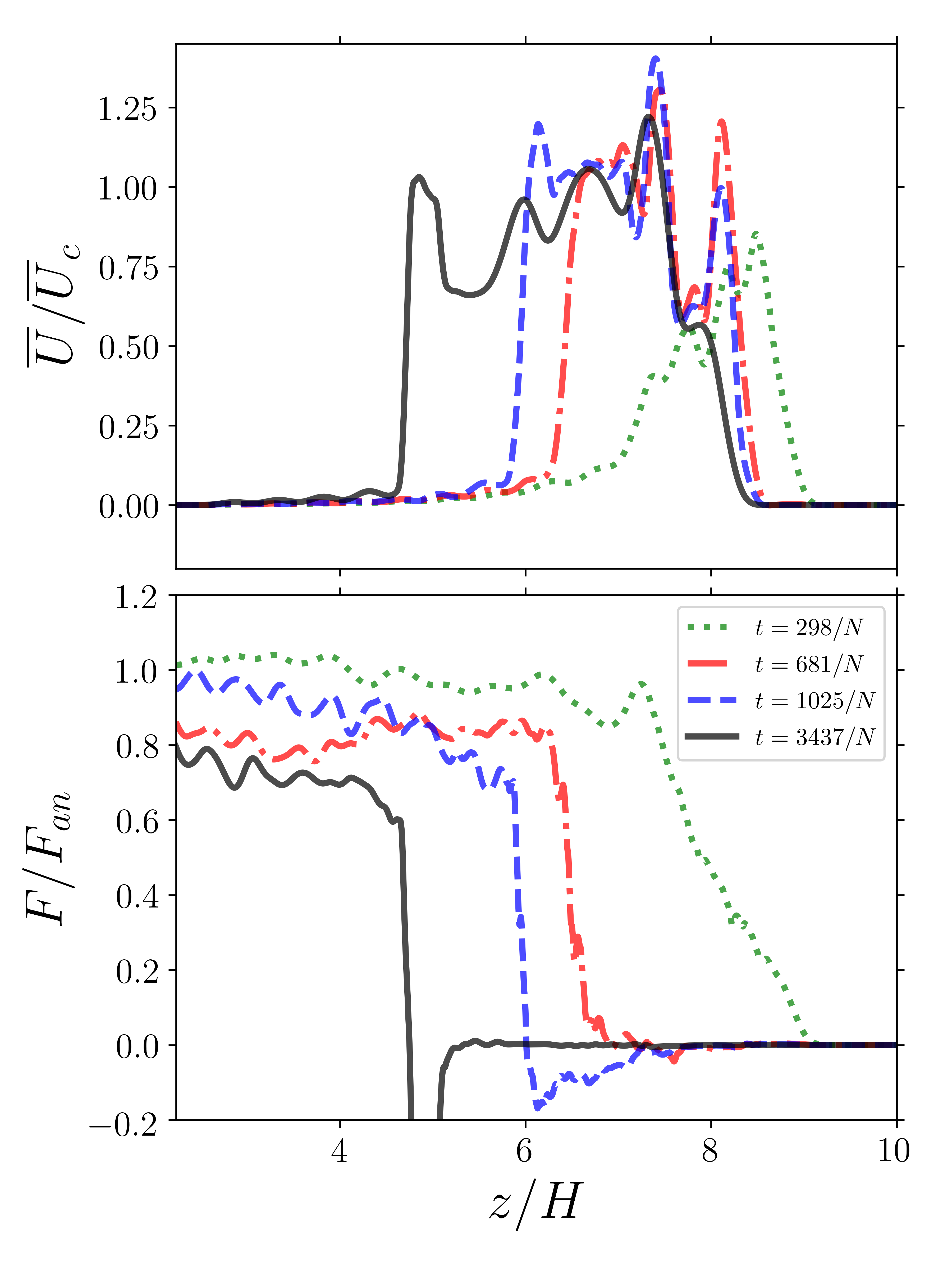}
    \caption{The mean horizontal flow velocity $\overline{U}(z, t)$
    (Eq.~\eqref{eq:mean_flow}) and the dimensionless momentum flux $F(z, t) /
    F_{an}$ (Eqs.~\eqref{eq:F_def} and~\eqref{eq:F_al}) in our fiducial
    simulation plotted at the same times as in Fig.~\ref{fig:snapshots}. The two
    distinct zones of mean flow are separated by a critical layer. The critical
    layer propagates toward lower $z$ due to momentum transport
    ($\pdil{F}{z}$).}\label{fig:nl_fluxes}
\end{figure}

In Fig.~\ref{fig:snapshots}, we present snapshots of $u_x$ and $\Upsilon$ at
various phases of the simulation. Note that $\Upsilon \lesssim
0.1$, so the density stratification does not deviate significantly from
equilibrium. The flow evolves through several distinct stages:
\begin{enumerate}
    \item At early times (top left panel), the flow resembles a linear IGW lower
        in the simulation domain but breaks down into smaller-scale features at
        higher $z$. Some characteristic swirling motion can be seen in the
        advected scalar $\Upsilon$, indicating Kelvin-Helmholtz instabilities.

    \item At a slightly later time (top right panel), the mean flow in
        $u_x$ becomes much more prominent and the critical layer $z_c$ has
        become much more definite. Small-scale fluctuations are still present in
        $u_x$ but at smaller amplitudes due to being in a denser region of the
        fluid.

    \item In the bottom left panel, the critical layer transition becomes very
        sharp, and small swirls of limited vertical extent in $\Upsilon$ at the
        location of the critical layer suggest that the Kelvin-Helmholtz
        instability is responsible for regulating the width of this transition.
        More discussion can be found in Section~\ref{ss:khi}.

    \item At the end of the simulation (bottom right panel), the
        critical layer has advanced downwards, but otherwise
        the flow shows very few significant qualitative differences from the
        previous snapshot. This suggests that the latter phase of the simulation
        has reached a steady state. Notably, the horizontal
        banded structure of $u_x$ in the upper, synchronized fluid does not
        continue to evolve (also visible in the top panel of
        Fig.~\ref{fig:nl_fluxes}), suggesting that momentum redistribution and
        mixing within the synchronized fluid are negligible.
\end{enumerate}

In Fig.~\ref{fig:nl_fluxes}, we plot the mean horizontal flow velocity
$\overline{U}$ (Eq.~\eqref{eq:mean_flow}) and the dimensionless momentum flux $F
/ F_{an}$ (Eqs.~\eqref{eq:F_def} and~\eqref{eq:F_al}) as a function of $z$ at
the times depicted in Fig.~\ref{fig:snapshots}. At each time, $\overline{U}$ is
close to zero below the critical layer, but then sharply increases to
$\overline{U}_c$ at the critical layer (i.e.\ the flow is ``spun-up''). Above
the critical layer, $\overline{U}$ varies slightly due to momentum transport
within the spun-up layer. This agrees with the expectation discussed in
Section~\ref{ss:crit_layer}.

Similarly, $F \lesssim F_{an}$ below the critical layer, and then decreases to
about zero above the critical layer. However, two notable deviations from the
discussion in Section~\ref{ss:crit_layer} can be observed: (i) the incident flux
on the critical layer fluctuates somewhat temporally, and (ii) there is a small
negative flux just above the critical layer at later times. These are addressed
in subsequent sections.

\subsection{Kelvin-Helmholtz Instability and Critical Layer Width}\label{ss:khi}

The formation of the critical layer is associated with a strong shear flow. What
is the width of this layer? Inspection of Fig.~\ref{fig:snapshots} suggests the
presence of the Kelvin-Helmholtz Instability (KHI) in the
critical layer. In a stratified medium, KHI occurs when the Richardson number
(Eq.~\eqref{eq:ri_def}) satisfies $\mathrm{Ri} \lesssim 1/4$
\citep[e.g.][]{shu1991physics}. It is natural to suspect that the shear flow
cannot steepen further than the onset of KHI\@. To test this, we compute the
local $\mathrm{Ri}$ for the shear flow around the critical layer.

\begin{figure}
    \centering
    \includegraphics[width=0.9\columnwidth]{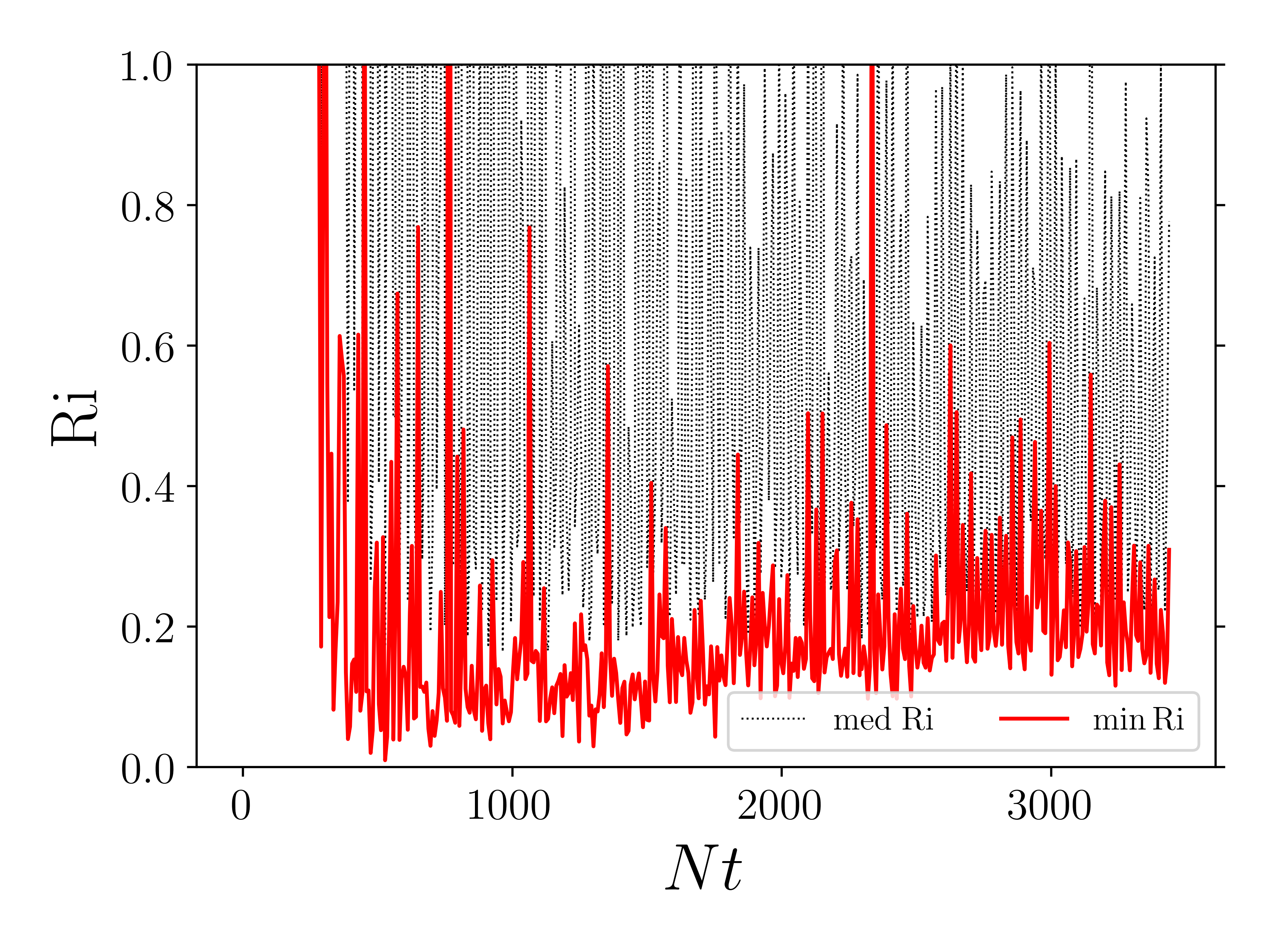}
    \caption{Local Richardson number (Eq.~\eqref{eq:ri_med_def}) of the flow at
    the critical layer over time (in units of $N^{-1}$) in our fiducial
    simulation. The solid red and dotted black lines denote respectively the
    minimum and median of $\mathrm{Ri}_x(x, t)$. These numbers measure the mean
    and spread in width of the critical layer over $x$. Note that $\mathrm{Ri}
    \sim \frac{1}{4}$ corresponds to the KHI, so this plot suggests the shear at
    the critical layer does not steepen past the onset of the
    KHI.}\label{fig:nl_f_ri}
\end{figure}

It is difficult to accurately measure the Richardson number, as it depends on
the derivative of the velocity. We measure $\mathrm{Ri}$ as follows: we first
assign an $\mathrm{Ri}_x(x, t)$ for every $x$ in the critical layer, then take
the median of $\mathrm{Ri}$ for the entire layer. $\mathrm{Ri}_x$ is computed
using the vertical distance over which the local $u_x$ increases from
$0.3\bar{U}_c$ to $\bar{U}_c$ (see Eq.~\eqref{eq:u_crit}). The value $0.3$ is
necessary to exclude the small mean flow generated in the weakly nonlinear
regime far below the
critical layer. This procedure can be written:
\begin{align}
    z_{CL, \min}(x, t) &\equiv \min
        \z{z\mid u_x(x, z, t) > 0.3\overline{U}_c},\\
    z_{CL, \max}(x, t) &\equiv \max
        \z{z\mid u_x(x, z, t) < \overline{U}_c},\\
    \mathrm{Ri}_x(x, t) &\equiv
        \p{\frac{N^2 \p{z_{CL, \max} - z_{CL, \min}}^2}{(0.7
            \overline{U}_c)^2}},\\
    \mathrm{Ri}(t) &\equiv \med_x\mathrm{Ri}\p{x, t}.\label{eq:ri_med_def}
\end{align}
We use the background buoyancy frequency to compute $\mathrm{Ri}$, as
fluctuations do not change $N^2$ significantly ($\sim 1\%$). To understand the
variation in $\mathrm{Ri}$ over $x$, we also compute $\min\limits_x
\mathrm{Ri}_x(x, t)$ (the maximum is very noisy). Both are shown in
Fig.~\ref{fig:nl_f_ri}. Absorption of incident IGWs quickly
decreases the Richardson number to between $0.25$ and $0.5$, characteristic of
the onset of the KHI\@.

This result suggests that the critical layer width is regulated by the
competition between steepening induced by IGW breaking and broadening due to
shear instability. This width does not vary significantly with resolution in our
resolved simulations (see Fig.~\ref{fig:agg}). As such, $\mathrm{Ri} \sim 0.5$
can be used to calculate the critical layer width in stars, where $N^2$ and
$\overline{U}_c$ (corresponding to the tidal frequency) are known.

\subsection{Flux Budget}\label{ss:flux_budget}

The downward propagation of the critical layer location $z_c(t)$ is driven by
the absorption of horizontal momentum flux at $z_c$, following
Eq.~\eqref{eq:zc_anal}. The flux budget at the critical layer can be decomposed
as
\begin{equation}
    F_i(t) = F_a(t) + F_r(t) + F_s(t),\label{eq:f_budget}
\end{equation}
where $F_i$ is the incident flux, $F_a$ is the absorbed flux, $F_r$ is the
reflected flux, and $F_s$ is some ``redistribution'' flux above the critical
layer, responsible for momentum redistribution within the synchronized upper
layer. Careful accounting of $F_s$ turns out to be important to obtain the
correct $F_a$ and resulting critical layer propagation. A more specific physical
interpretation of $F_s$ is unclear; it is somewhat tempting but unfounded to
identify $F_s$ with the transmitted flux. In these simulations, we find $F_s <
0$, corresponding to net momentum transport \emph{into} the critical layer from
the synchronized layer above it.

\begin{figure}
    \centering
    \includegraphics[width=0.9\columnwidth]{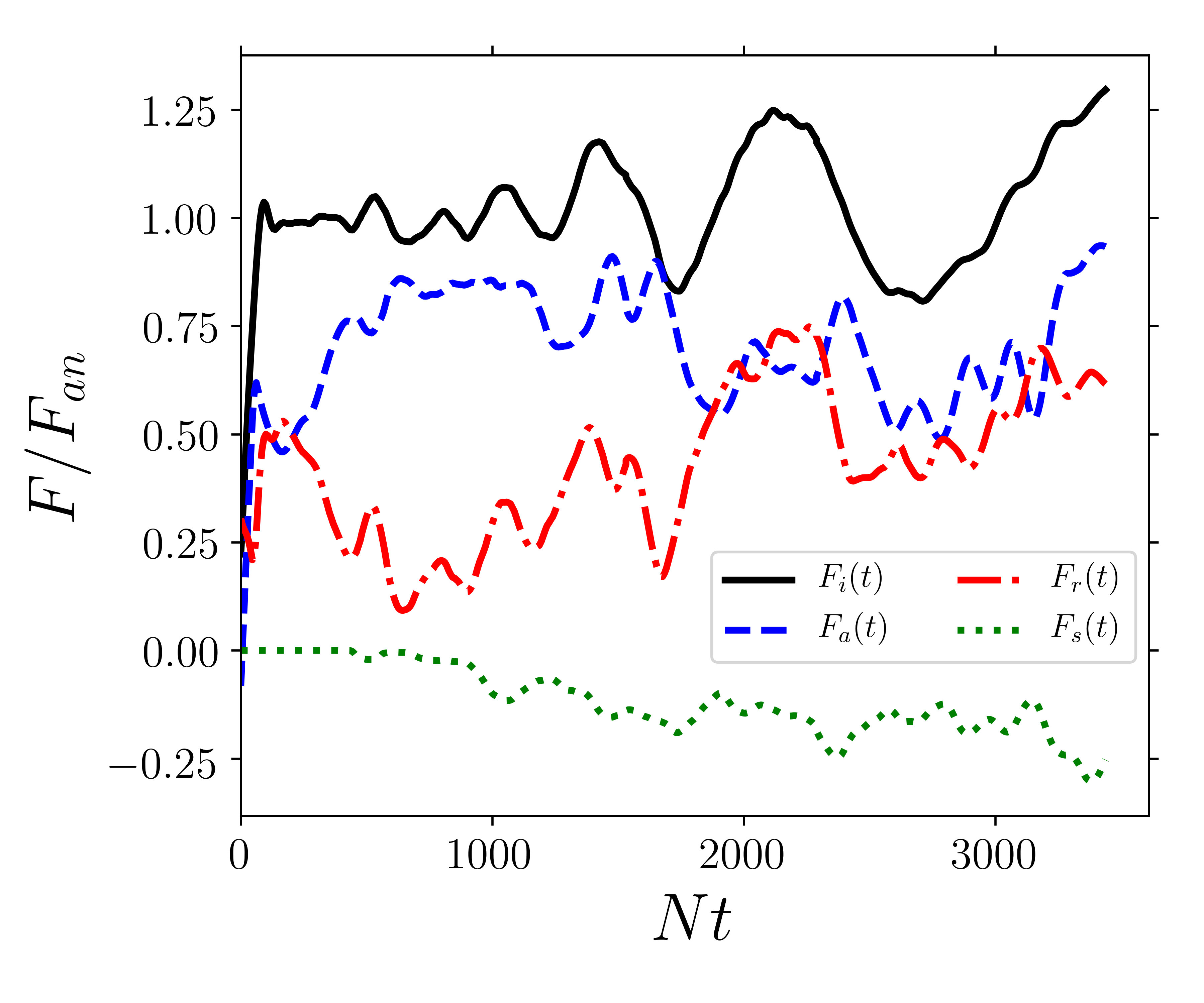}
    \caption{Momentum flux decomposition calculated from the simulation. Plotted
    are the four components of the horizontal momentum flux budget over time
    (see Eq.~\eqref{eq:f_budget}), in units of the analytical estimate for the
    incident wave flux $F_{an}$ (Eq.~\eqref{eq:S_lin}): $F_i$, the flux incident
    on the critical layer; $F_a$, the flux absorbed by the critical layer;
    $F_r$, the flux reflected at the critical layer; and $F_s$, the flux inside
    the synchronized layer.}\label{fig:nl_f_amps2}
\end{figure}

After measuring $z_c$ (see Section~\ref{ss:cl_prop}) and $F(z)$
(Eq.~\eqref{eq:F_def}) at each time step, we determine each of $F_i$, $F_a$,
$F_r$, $F_s$ as follows:
\begin{align}
    F_i(t) &= F_{an}A_i^2(t),\\
    F_r(t) &= F_i(t) - \frac{1}{H}
        \int\limits_{z_c - \Delta z - H}^{z_c - \Delta z}F(z, t)\;\mathrm{d}z
        ,\label{eq:fr_def}\\
    F_s(t) &=
        \frac{1}{\Delta z}\int\limits_{z_c}^{z_c + \Delta z}
            \min(F(z, t), 0)\;\mathrm{d}z,
        \label{eq:fs_def}\\
    F_a(t) &= F_i - F_r - F_s.\label{eq:fa_def}
\end{align}
Fig.~\ref{fig:nl_f_amps2} depicts the four components of this flux
decomposition.
Below the critical layer, we average over an interval of length $H$, also the
vertical wavelength. The offset $\Delta z$ is necessary to make the measurement
of the incident flux unaffected by the turbulence within the critical layer
itself. The width of the critical layer is limited by $\mathrm{Ri} \lesssim 1$
(see Section~\ref{ss:khi}), which bounds its vertical extent $\sim
\frac{1}{\abs{k_{z}}}$. We empirically found an offset of $\Delta z =
\frac{3}{\abs{k_z}}$ was necessary to be sufficiently far from strong
fluctuations near the critical layer.

Above the critical layer, we observe that the $F_s$ feature has varying width
(compare e.g.\ the $t = 1171.4/N$ and $t = 3437.8/N$ lines in the bottom panel
of Fig.~\ref{fig:nl_fluxes}) but contributes significantly to the total
flux budget. We average only where $F < 0$ so that $F_s$ is robust to such width
variations. We find that this is an accurate way of measuring $F_s$ and
determining $F_a$.

\subsection{Critical Layer Propagation}\label{ss:cl_prop}

With a careful determination of $F_a$, we can make predictions for the
propagation of $z_c(t)$ and compare to the measured propagation in the
simulation. In principle, $z_c$ is the location where the incident flux
significantly attenuates. In the simulation, shear turbulence causes $F$ to have
significant spatial and temporal fluctuations that translate to large temporal
fluctuations in $z_c(t)$. To minimize these spurious fluctuations, we measure
the location of the critical layer using a spatial average of where flux
deposition occurs:
\begin{align}
    z_{c, \min}(t) &\equiv \min_z \z{z: F(z, t) > 0.3F_{an}},\\
    z_{c, \max}(t) &\equiv \max_z \z{z: F(z, t) < 0.3F_{an}},\\
    z_c(t) &\equiv \frac{z_{c, \min}(t) + z_{c, \max}(t)}{2}.\label{eq:zc_def}
\end{align}
Measuring $z_c$ in other ways does not significantly change the results of the
analysis.

\begin{figure}
    \centering
    \includegraphics[width=0.9\columnwidth]{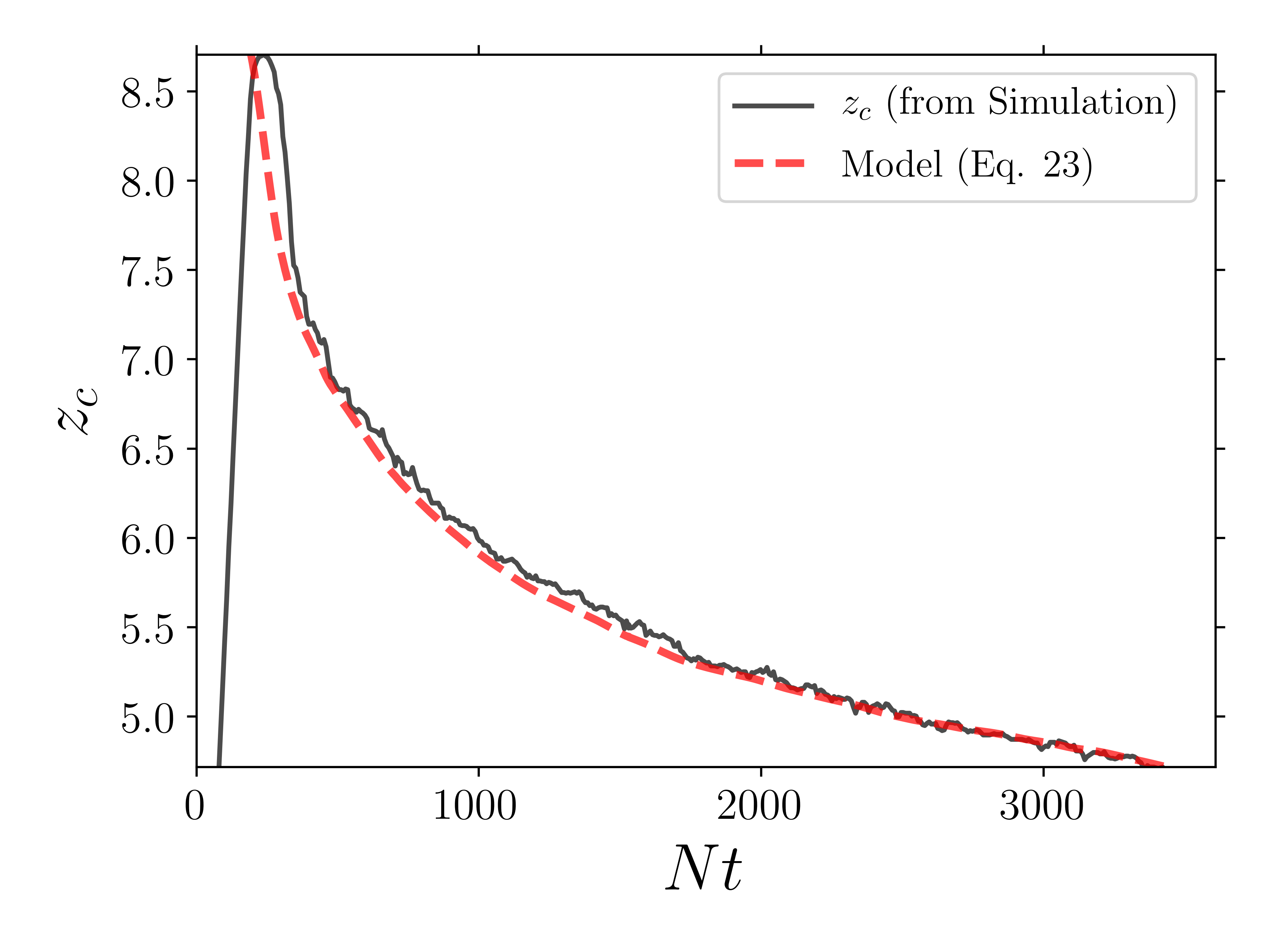}
    \caption{Propagation of the critical layer over time. Shown are: (solid black)
    $z_c(t)$ from simulation data, and (dashed red) model for $z_c(t)$ using direct
    integration of Eq.~\eqref{eq:zc_anal} for $F_a(t)$ measured from simulation
    data (described in Eq.~\eqref{eq:fa_def}). The model uses the end of the
    simulation as its initial condition and integrates backwards, as the
    critical layer is still forming at earlier times. The agreement of the model
    with the simulation shows Eq.~\eqref{eq:zc_anal} is a good description of
    the evolution of $z_c$.}\label{fig:nl_front}
\end{figure}

In Fig.~\ref{fig:nl_front} we plot the numerically measured $z_c$ against
numerical integration of Eq.~\eqref{eq:zc_anal} using the measured $F_a(t)$.
Since the critical layer is still forming at early times, we solve
Eq.~\eqref{eq:zc_anal} by integrating backwards from the end of the simulation
($t = t_f$), using $z_c(t_f)$ as the initial condition. From
Fig.~\ref{fig:nl_front}, we see that the agreement between the measured $z_c(t)$
and its estimate via $F_a(t)$ is excellent.

By time-averaging the numerically measured $F_a$, we find $\ev{F_a}_t \approx
0.71F_{an}$. Note that $F_a < F_{an}$, so momentum flux absorption at the
critical layer is incomplete. This is due to reflection of waves off the
critical layer, which carry momentum downward.

\subsection{Non-absorption at Critical Layer}\label{ss:reflectivity}

To further understand the behavior at the critical layer, we compare two
reflective behaviors observed in the simulation: (i) the presence of a reflected
wave with the same frequency as the incident wave (i.e.\ with wave vector
$\bm{k}_r = k_{x}\uv{x} - k_{z}\uv{z}$), and (ii) the reflected flux $F_r$. The
reflected wave amplitude and flux need not agree exactly if some reflected flux
is in higher-order modes, which is indeed the case in our simulations. Both are
of physical interest, however: the reflected wave amplitude is essential for
setting up standing modes in a realistic star, while the flux is important for
accurately tracking angular momentum transfer during synchronization.

To measure the reflected wave amplitude $A_r(t)$, we use an approach similar to
the calculation of $A_i(t)$ (Eq.~\eqref{eq:ahat_def}):
\begin{equation}
    A_r(t) = \max_{\delta x}\frac{\int\limits_{z_b}^{z_t}\int\limits_0^{L_x}
        \overline{\rho}\p{\bm{u} \cdot \at{\bm{u}_{an,
        \bm{k}_r}}_{x = x + \delta x}}\;\mathrm{d}x\mathrm{d}z}
        {\int\limits_{z_b}^{z_t}\int\limits_0^{L_x}
        \overline{\rho}\abs{\bm{u}_{an}}^2\;\mathrm{d}x\mathrm{d}z},
        \label{eq:ar_def}
\end{equation}
where $z_b = z_0 + 3\sigma$ and $z_t = z_b + H$ as before. The primary
difference from Eq.~\eqref{eq:ahat_def} is the introduction of free parameter
$\delta x$, the horizontal phase offset of the reflected wave. Since $\delta x$
is unknown \emph{a priori}, we choose $\delta x \in [0, 2\pi]$ that maximizes
$A_r(t)$. In our simulation, the phase offset $\phi_r(t) \equiv k_x \delta x(t)
$ is consistent with reflection off a moving boundary at $z_c$,
i.e.\ $\abs{\pdil{\phi_r}{t}} \simeq 2\abs{\pdil{(k_{z}z_c)}{t}}$.

\begin{figure}
    \centering
    \includegraphics[width=0.9\columnwidth]{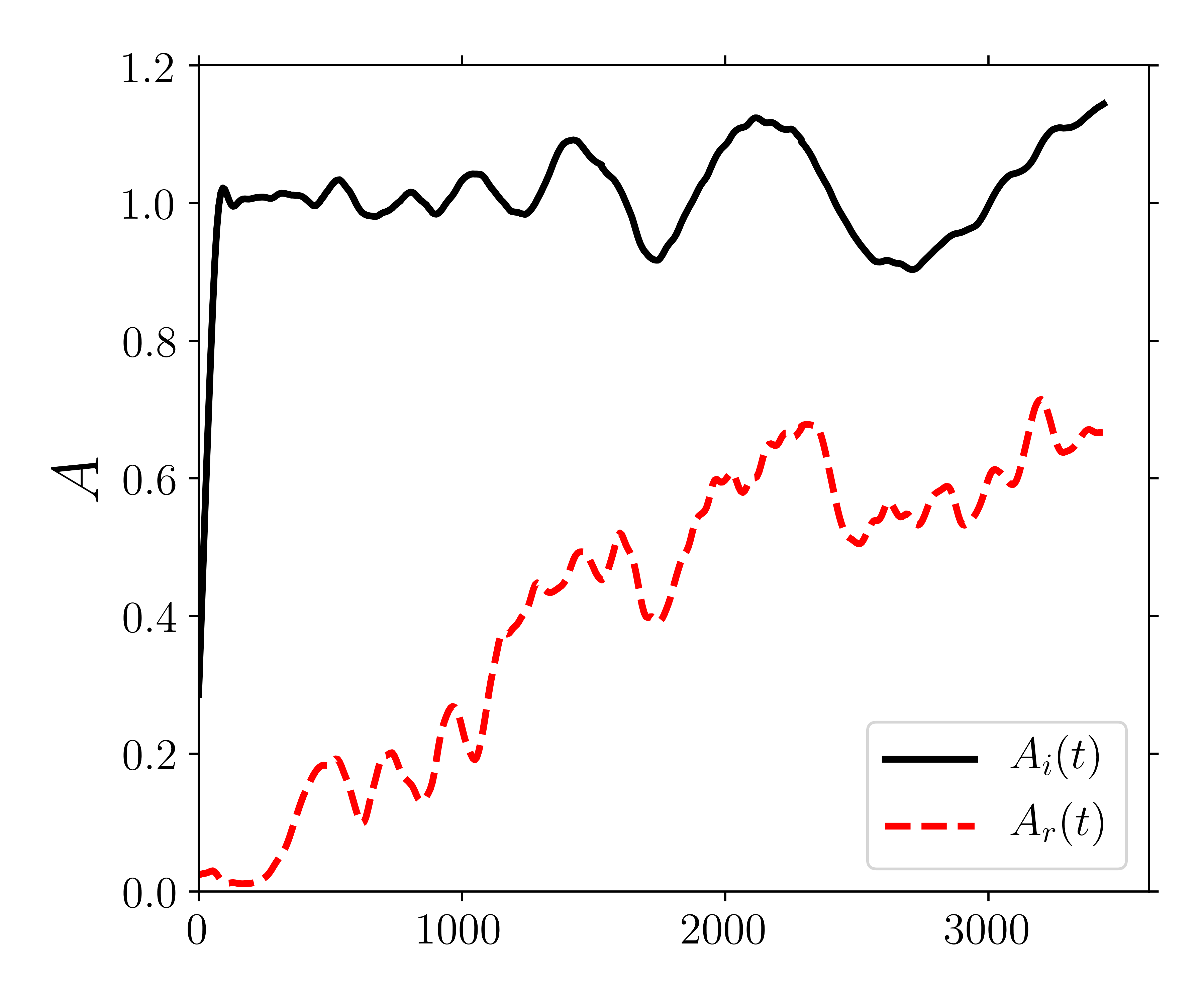}
    \caption{The incident wave amplitude $A_i(t)$ (solid black) and the
    reflected wave amplitude $A_r(t)$ (dashed red) just above the forcing
    zone.}\label{fig:nl_f_amps}
\end{figure}

Fig.~\ref{fig:nl_f_amps} illustrates the behaviors of $A_i$ and $A_r$. Both vary
significantly in time but their mean values appear to converge towards the end
of the simulation.

Since $A_i(t), A_r(t)$ vary somewhat over time, we perform time averaging over
interval of four wave periods, denoted by angle brackets. We can then define the
amplitude reflectivity
\begin{equation}
    \mathcal{R}_A(t) \equiv \frac{\ev{A_r}(t)}{\ev{A_i}(t)}
        .\label{eq:Ra_def}
\end{equation}

\begin{figure}
    \centering
    \includegraphics[width=0.9\columnwidth]{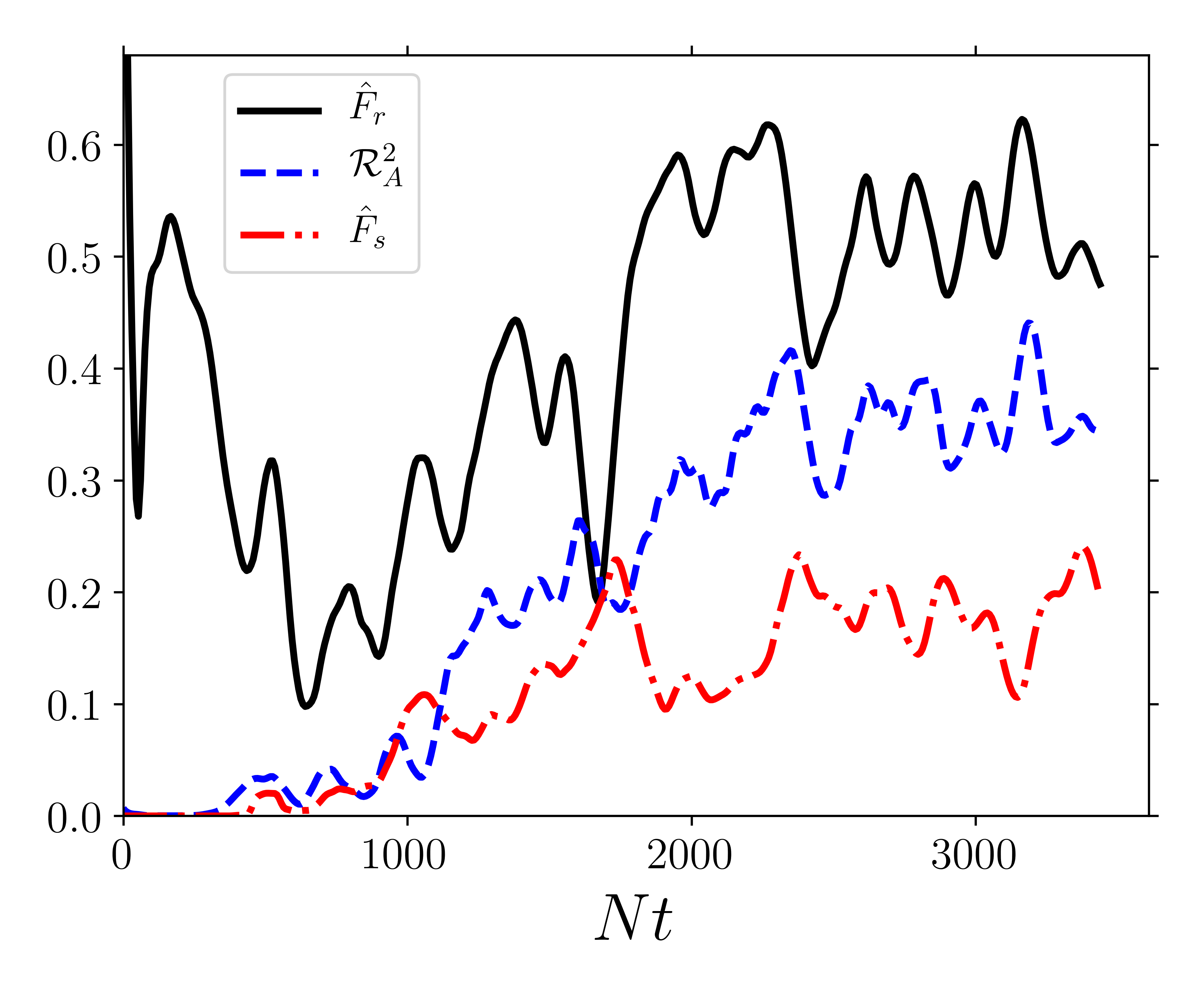}
    \caption{$\mathcal{R}_A^2$, $\hat{F}_r$, and redistribution flux $\hat{F}_s$
    [Eqs.~(\ref{eq:Ra_def}--\ref{eq:srefl_def})] as a function of time (in
    units of $N^{-1}$). These quantities seem to become comparatively stable
    past about $t = 2500/N$, indicating that an asymptotic value may have been
    reached. That $\hat{F}_r \gtrsim \mathcal{R}_A^2$ implies a substantial
    fraction of reflected flux is in higher-order modes than the reflected
    IGW.}\label{fig:nl_f_refl}
\end{figure}

We compare the square of the reflectivity to the ratios of $F_r$ and $-F_s$ to
$F_i$, as $F \propto A^2$ (Eq.~\eqref{eq:S_lin}). We define
\begin{align}
    \hat{F}_r &\equiv \frac{\ev{F_r}(t)}{\ev{F_i}(t)}, \label{eq:srefl_def1}\\
    \hat{F}_s &\equiv -\frac{\ev{F_s}(t)}{\ev{F_i}(t)}. \label{eq:srefl_def}
\end{align}
Fig.~\ref{fig:nl_f_refl} shows $\mathcal{R}_A^2$, $\hat{F}_r$, and $\hat{F}_s$
as functions of time. The three quantities appear to be roughly stationary for
$t \gtrsim 2500/N$. Modest fluctuations ($\sim 20\%$) in $A_i$ do not affect our
reflectivity results thanks to the time averaging used in
Eqs.~\eqref{eq:Ra_def}--\eqref{eq:srefl_def}.
We see that in general $\hat{F}_r \gtrsim \mathcal{R}_A^2$, conforming with the
expectation that the reflected flux consists of the simple reflected mode and
higher order modes as well.

\subsection{Resolution Study}\label{ss:convergence}

Although throughout this paper we focused on our fiducial simulation with
$\mathrm{Re} = 1024$ and resolution $N_x=768$, $N_z=3072$, we also ran a suite
of simulations varying the resolution and corresponding Reynolds number
(Tab.~\ref{tab:params}). We find that our global, quantitative measurements in
the simulations ($\mathcal{R}_A^2$, $F_r$, $F_s$, and $\mathrm{Ri}$) are very
similar for our highest Reynolds numbers (1024 and 2048).

\begin{figure}
    \centering
    \includegraphics[width=0.9\columnwidth]{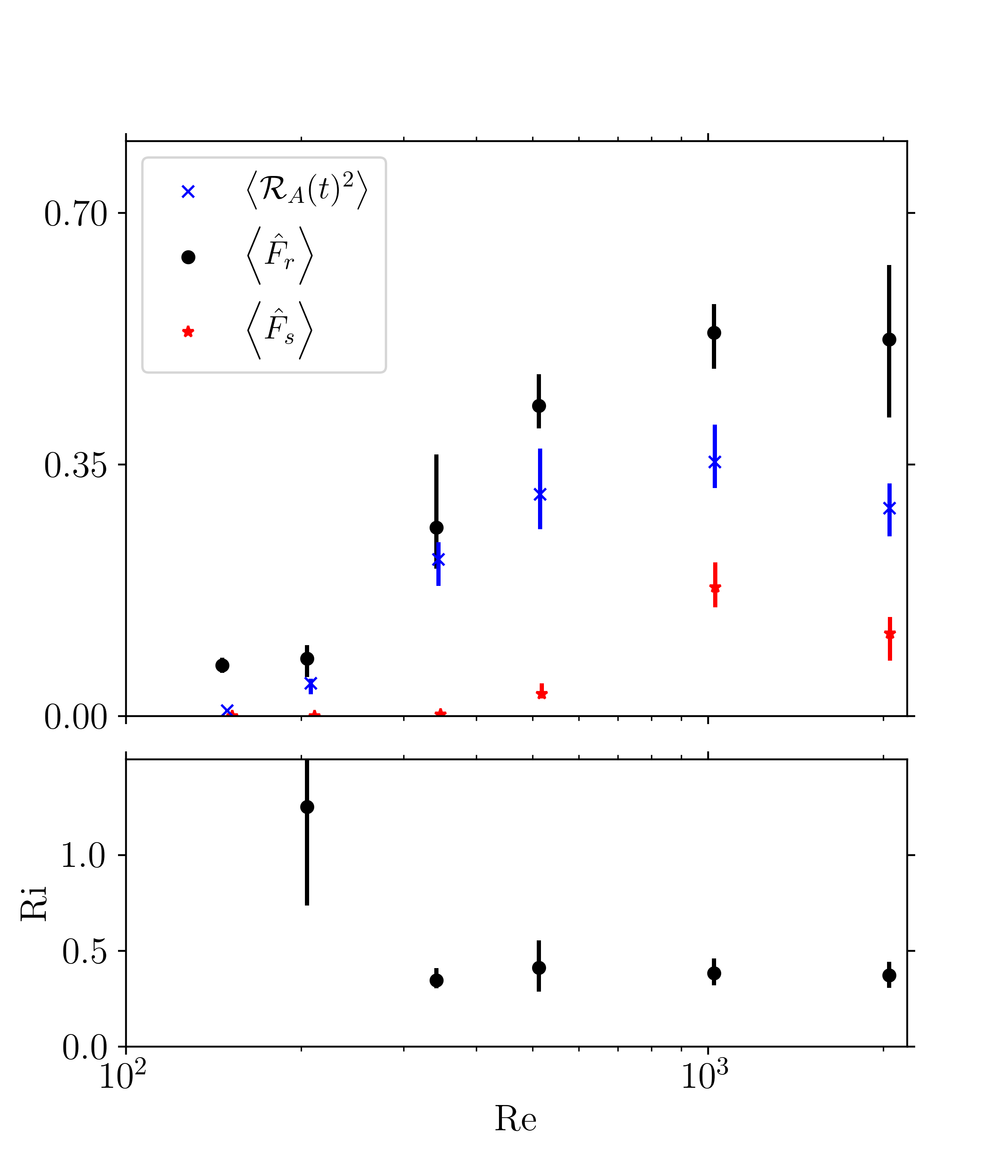}
    \caption{Convergence of the median $\hat{F}_r$, $\mathcal{R}_A^2$,
    $\hat{F}_s$, and $\mathrm{Ri}$ (Eqs.~\eqref{eq:Ra_def}--\eqref{eq:srefl_def}
    and~\eqref{eq:ri_def} respectively) in simulations with varying resolution
    and viscosity as given in Tab.~\ref{tab:params}. Vertical bars show the
    temporal variation of each measurement between the $16\%$ and $84\%$ range.
    Small horizontal displacements are made for data points at identical
    $\mathrm{Re}$ for readability. Note that simulations with larger
    $\mathrm{Re}$ correspond to smaller viscosity and are more physically
    realistic. At the smallest $\mathrm{Re}$ value, $\mathrm{Ri} \approx 50$ is
    too large to fit on the plot.}\label{fig:agg}
\end{figure}

For each simulation in Tab.~\ref{tab:params}, we compute the median values of
$\mathcal{R}_A^2$, $\hat{F}_r$, $\hat{F}_s$, and $\mathrm{Ri}$
[Eqs.~(\ref{eq:Ra_def}--\ref{eq:srefl_def}) and~\eqref{eq:ri_def}
respectively] over the last $1/4$ of the simulation time, when these quantities
have reached their asymptotic values. These results are shown in
Fig.~\ref{fig:agg}.

As the simulation resolution increases and the viscosity decreases, we find that
the Richardson number decreases, while the reflection and redistribution fluxes
increase. The Richardson number is roughly constant for $\mathrm{Re} > 200$ with
a value of $\mathrm{Ri} \sim 0.4$. The behavior of the fluxes is more
complicated. While the fraction of reflected and redistributed flux is similar
for our simulations with $\mathrm{Re} = 1024$ and $2048$, higher resolution
simulations would be required to determine these flux fractions in the limit
$\mathrm{Re}\rightarrow\infty$.

Nevertheless, the difference in behavior of $\mathrm{Ri}$ and the flux
reflectivity as $\mathrm{Re}$ is varied is in tension with
Eq.~\eqref{eq:crit_coeffs}. This tension is natural: Eq.~\eqref{eq:crit_coeffs}
is derived from a linear theory, while fluid motion within the critical layer is
turbulent, so reflection at the critical layer cannot be captured by the linear
theory.

\section{Summary and Discussion}\label{s:discussion}

\subsection{Key Results}

In this paper, we have performed numerical simulations of nonlinear breaking of
IGWs in a stratified isothermal atmosphere. Such a setup represents the
plane-parallel idealization of the outer stellar envelope. Our simulations use
the spectral code Dedalus \citep{dedalus,dedalus2}, and are carried out in 2D.
We observe spontaneous formation of a critical layer that separates a
``synchronized'' upper layer of fluid and a lower layer with no mean
horizontal flow. This critical layer then propagates downwards as incident IGWs
break and deposit horizontal momentum to the fluid (see Fig.~\ref{fig:snapshots}
for snapshots from our fiducial simulation). Our primary conclusions regarding
the evolution of the critical layer are as follows:
\begin{enumerate}
    \item The width of the turbulent critical layer is determined by requiring
        the local Richardson number (Eq.~\eqref{eq:ri_def}) $\mathrm{Ri} \sim
        0.5$ (see Fig.~\ref{fig:nl_f_ri}).

    \item The location of the critical layer $z_c(t)$ can be predicted by
        careful measurement of the absorbed horizontal momentum flux at the
        critical layer (see Eq.~\eqref{eq:zc_anal} and
        Fig.~\ref{fig:nl_front}).

    \item The absorption of IGW momentum flux at the critical layer is
        incomplete. The critical layer only absorbs $\sim 70\%$ of the incident
        flux in our highest resolution simulations (see Fig.~\ref{fig:agg}). The
        reflected flux is carried away from the critical layer as both
        lowest-order reflected waves and waves with larger $z$ wavenumbers.
\end{enumerate}

\subsection{Discussion}

In this paper, we have studied the nonlinear behavior of IGWs with $\abs{k_x /
k_z} \sim 1 / (2\pi)$ in a plane-parallel geometry. Tidally excited IGWs in
binary stars have horizontal wavenumber $k_\perp \sim 1/R$ (where $R$ is the
stellar radius) much smaller than the radial wavenumber $k_r$. While our
simulations do not satisfy $\abs{k_x/k_z}\ll 1$, the qualitative behavior is likely
to be similar, as the turbulence driving the critical layer dynamics occurs at
scales significantly smaller than either $1/k_x$ or $1/\abs{k_z}$. Simulating
IGWs with $k_x \ll \abs{k_z}$ is more challenging numerically and we defer its
exploration to future work.

It is interesting to compare our work with that of \citet{barker_ogilvie}, who
studied inward-propagating IGWs in solar-type stars and their nonlinear breaking
due to geometric focusing. In their numerical simulations in a 2D polar
geometry, they found no evidence for reflected waves, contrary to our result.
Note that their simulations were run with substantially higher viscosity, or
lower resolution, than explored here, and their effective Reynolds number (equal
to $1/\lambda$ in their notation) is of order $10$. We also find at low Reynolds
numbers that there is negligible wave reflection.

Regardless of the limitations inherent in our simulations (e.g.\ plane-parallel
geometry), our results shed light on the physical mechanism of tidal heating in
close binaries. In particular, our simulations indicate that energy dissipation
occurs in a narrow critical layer. The star heats up from outside-in as the
critical layer propagates inwards. This tidal heating profile differs flom that
used by \citet{tidal_novae}. We plan to study this issue in a future work.

\section{Acknowledgements}\label{s:ack}

This work has been supported in part by the NSF grant AST-17152. YS is supported
by the NASA FINESST grant 19-ASTRO19-0041.
DL is supported by the Princeton Center for Theoretical Sciences and Lyman
Spitzer Jr fellowships. Computations were conducted with support by the NASA
High End Computing (HEC) Program through the NASA Advanced Supercomputing (NAS)
Division at Ames Research Center on Pleiades with allocation GID s1647.

\bibliographystyle{mnras}
\bibliography{Su_IGW_break}

\clearpage
\onecolumn
\appendix

\section{Derivation of Fluid Equations}\label{s:equation_deriv}

We aim to model wave dynamics over multiple density and pressure scaleheights.
Start with the compressible Euler equations.
\begin{align}
    \partial_t \rho + \p{\bm{u}\cdot\bm{\nabla}} \rho
            + \rho \bm{\nabla}\cdot\bm{u} =\, &0, \\
    \partial_t s + \p{\bm{u}\cdot\bm{\nabla}} s =\, &0, \\
    \partial_t \bm{u} + \p{\bm{u}\cdot\bm{\nabla}}\bm{u} +
        \frac{1}{\rho}\bm{\nabla} p =& \, \bm{g},
\end{align}
where $s$ is the entropy. The pressure is calculated from the entropy and
density using the equation of state
\begin{equation}
    \frac{s}{c_p} = \log \frac{p^{1/\gamma}}{\rho},
\end{equation}
where $c_p$ is the specific heat at constant pressure, and $\gamma$ is the ratio
of specific heats. For computational ease, it is convenient to filter out the
fast sound waves from these equations. One approach is to assume pressure
perturbations are small \citep[yielding the ``pseudo-incompressible''
equations,][]{anel_part2}, or that all thermodynamic perturbations are small
\citep[yielding the ``anelastic'' equations,][]{anel_part1}. In these
approximations, one of the thermodynamic equations is replaced by a constraint
equation: $\bm{\nabla}\cdot \p{p_0^{1/\gamma} \bm{u}}=0$ for
pseudo-incompressible; $\bm{\nabla}\cdot\p{\rho_0 \bm{u}} =0$ for anelastic.
Here $p_0$ and $\rho_0$ are the background density and pressure profiles. The
pressure in the momentum equation can be interpreted as a Lagrange multiplier
which enforces the constraint \citep{anel_part2}. Upon linearization, both
approximations conserve a wave energy
\begin{equation}
    E_w = \frac{1}{2}\rho_0 \abs{\bm{u}}^2
            + \frac{1}{2}\frac{g^2}{N^2} \frac{(\rho')^2}{\rho_0}
        = \frac{1}{2}\rho_0 \abs{\bm{u}}^2
            + \frac{1}{2}\rho_0 N^2 \frac{(s')^2}{\abs{\bm{\nabla} s_0}^2},
            \label{eq:ewave}
\end{equation}
where $\rho'$ and $s'$ represent the density and entropy perturbations.

Rather than assume thermodynamic perturbations are small, we instead filter out
sound waves by taking the limit $\gamma\rightarrow \infty$. Then the entropy and
log density are proportional to each other, and the entropy equation becomes
\begin{equation}
    \partial_t \log \rho + \p{\bm{u}\cdot \bm{\nabla}} \log \rho = 0.
\end{equation}
Together with mass conservation, this implies
\begin{equation}
    \bm{\nabla} \cdot \bm{u} = 0.
\end{equation}
We solve these equations together with the normal momentum equation.

Although these equations are non-standard, they have various desirable
properties. They conserve mass and momentum, and the linearized equations
conserve the wave energy $E_w$ [Eq.~\eqref{eq:ewave} above] similar to the
pseudo-incompressible equations and anelastic equations. Our equations also
satisfy the same linear dispersion relation as the fully compressible equations
in the limit of large sound speed \citep[this is also true for the anelastic
equations, but not pseudo-incompressible,][]{anel_part1, anel_part2}. Thus, the
vertical propagation of internal gravity waves is similar to the
pseudo-incompressible equations and anelastic equations.

In this work we are interested in waves which reach large amplitudes and break.
For breaking waves, \citet{achatz_igw_pi} suggests that the anelastic equations
may miss important effects. Although the pseudo-incompressible equations may
capture wave-breaking more accurately, they are more complicated, and do not
satisfy the correct dispersion relation to order $(k_z H)^{-2}$
\citep{anel_part2}. In the absence of a clear choice to study this wave breaking
problem, we have elected to use these simple equations derived in the
$\gamma\rightarrow\infty$ limit.

\section{Forcing Solution}\label{s:force_solved}

To solve for the linear excited IGW amplitude due to bulk forcing (see
Eq.~\eqref{eq:vol_drive}), we consider the linearized system of equations, with
all dynamical variables having dependence $u_z(x, z, t) = \tilde{u_z}(z)
e^{ik_xx - i\omega t}$. Thus, $\pdil{}{t} \to -i\omega t$, $\pd{}{x} \to ik_x$,
and the dynamical fluid
equations become (see Eqs.~\eqref{se:nl_orig} and~\eqref{se:nl_upsilon}):
\begin{align*}
    \rd{u_{z}}{z} + ik_xu_x &= 0,\\
    -i\omega u_x + ik_x \varpi + gHik_x \Upsilon &= 0,\\
    -i\omega u_{z} + \rd{\varpi}{z} + gH\rd{\Upsilon}{z}
        - \frac{\varpi}{H} &= 0,\\
    -i\omega \Upsilon - \frac{u_{z}}{H} &=
        C\exp\s{-\frac{(z - z_0)^2}{2\sigma^2}} \equiv \mathcal{C}(z).
\end{align*}
These can be recast solely in terms of $u_{z}$ as
\begin{align*}
     \rtd{u_{z}}{z} - k_x^2u_{z} - \frac{1}{H}\rd{u_{z}}{z}
        + u_{z}\frac{N^2k_x^2}{\omega^2} &=
    -\frac{gk_x^2}{\omega^2}\mathcal{C}(z).
        .\label{eq:narrow_inhomo}
\end{align*}
The homogeneous solutions are of form $u_{z,\pm}(z) = \exp\s{\p{\frac{1}{2H} \pm
ik_z}\p{z - z_0}}$ where $k_z$ satisfies the dispersion relation
(Eq.~\eqref{eq:disp_rel}). We compute the solution to the inhomogeneous ODE by
the method of variation of parameters. The Wronskian is
\begin{equation}
    W \equiv \det \begin{vmatrix}
        u_{z,+} & u_{z,-} \\[4pt]
        \rdil{u_{z,+}}{z} & \rdil{u_{z,-}}{z}
    \end{vmatrix} = -2ik_ze^{z/H}.
\end{equation}
The general solution is then
\begin{equation}
    u_z = -u_{z,+}\int \frac{1}{W} u_{z,-} \p{-\frac{gk_x^2}{\omega^2}
            \mathcal{C}(z)}\;\mathrm{d}z
        + u_{z,-}\int \frac{1}{W} u_{z,+} \p{-\frac{gk_x^2}{\omega^2}
            \mathcal{C}(z)}\;\mathrm{d}z.
\end{equation}
Taking these integrals and applying the boundary conditions $u_z\p{z \to \infty}
= u_{z,+}$, $u_z\p{z \to -\infty} = u_{z,-}$ give the exact solution:
\begin{align}
    u(z) &= \frac{\sqrt{\pi}C\sigma}{2^{3/2}ik_z}\frac{gk_x^2}{\omega^2}
            \exp\s{\frac{\p{\frac{\sigma^2}{2H} \pm ik_z\sigma^2}^2}{2\sigma^2}}
                \s{-u_{z,+}\p{1 + \erf\p{\xi_+}}
                    + u_{z,-} \p{\erf\p{\xi_-} - 1}},\\
    \xi_{\pm} &\equiv \s{\frac{z - z_0}{\sqrt{2\sigma^2}} +
        \frac{\sigma}{2^{3/2}H} \pm \frac{ik_z\sigma}{\sqrt{2}}}
\end{align}
Here, the error function is defined $\erf(z) \equiv
\p{2/\sqrt{\pi}}\int\limits_0^z \exp\p{-t^2}\;\mathrm{d}t$. If we are concerned
with only $z$ scales significantly larger than $\sigma$, then we may take
$\erf(\xi_{\pm}) \approx \Theta(z - z_0)$ (the Heaviside step function). If we
further assume $\abs{k_zH} \gg 1$ and restore the $e^{ik_xx - i\omega t}$
factor, we recover Eq.~\eqref{eq:uz_lin} in the main text
\begin{equation}
    u_{z}(x, z, t) = -\frac{C}{2ik_z}\frac{gk_x^2}{\omega^2}
        e^{-k_z^2\sigma^2 / 2}
        \sqrt{2\pi \sigma^2} e^{ik_xx - i\omega t} \times
    \begin{cases}
        \exp\s{\p{\frac{1}{2H} + ik_z}\p{z - z_0} + i\frac{k_z\sigma^2}{2H}}
            & \text{for }z > z_0,\\[5pt]
        \exp\s{\p{\frac{1}{2H} - ik_z}\p{z - z_0} - i\frac{k_z\sigma^2}{2H}}
            & \text{for }z > z_0.
    \end{cases}
\end{equation}
Note that in the main text, this approximate form is used to compute
$\bm{u}_{an}$, as it is easier to work with and sufficiently accurate in the
regions of interest (many $\sigma$ away from $z_0$).

\section{Equation Implementation}\label{se:strat_impl}

The system of equations we wish to simulate consists of
Eqs.~\eqref{eq:nl_incomp},~\eqref{eq:nl_upsilon_u}, and~\eqref{eq:vol_drive}.
The nonlinear terms in the these equations will transfer energy from lower
wavenumbers to higher wavenumbers. Since spectral codes have no numerical
diffusion, explicit diffusion must be added. To ensure the non-ideal
system conserves horizontal momentum exactly, we begin by adding diffusion
terms to the flux-conservative form of the Euler fluid equations
(equivalent to Eqs.~\ref{se:nl_orig}):
\begin{subequations}
    \begin{align}
        \bm{\nabla} \cdot \bm{u} &= 0,\\
        \partial_t \rho + \bm{\nabla} \cdot (\rho \bm{u} - \nu
            \bm{\nabla}(\rho - \overline{\rho})) &= 0,\label{eq:visc_cons_mom}\\
        \partial_t (\rho \bm{u}) + \bm{\nabla} \cdot (\rho \bm{u} \bm{u}
            + \mathrm{diag}(\rho \varpi)
            - \nu \rho \bm{\nabla}\bm{u})
            + \rho g \uv{z} &= 0.
    \end{align}
\end{subequations}
For simplicity, we use the same diffusivity $\nu$ for both the momentum and mass
diffusivities. Although mass diffusivity is not physical, we include it for
numerical stability. We choose the mass diffusion term to conserve mass, and not
to affect the background density profile.

It is necessary to mask out nonlinear terms in the forcing zone using a form
similar to Eq.~\eqref{eq:Gamma}. In the absence of this mask, a nonphysical mean
flow localized to the forcing zone develops. We use the mask
\begin{equation}
    \Gamma_{NL}(z) = \frac{1}{2}\s{2
        + \tanh \frac{z - (z_0 + 8\sigma)}{\sigma}
        - \tanh \frac{z - z_B}{\sigma}}.
\end{equation}

Including the damping zones and forcing terms as described in
Section~\ref{s:numerics}, and again making change of variables to $\Upsilon,
\varpi$, we finally obtain the full system of equations as simulated in Dedalus:
\begin{subequations}\label{se:dedalus_eqs}
    \begin{align}
        \bm{\nabla} \cdot \bm{u} ={}& 0,\\
        \partial_t \Upsilon - \frac{u_z}{H}
            ={}& -\Gamma(z) \Upsilon
                + \frac{F}{\overline{\rho}(z)}e^{-\frac{(z - z_0)^2}{2\sigma^2}}
                    \cos \p{k_xx - \omega t}\nonumber\\
            & + \Gamma_{NL} \bigg[-\p{\bm{u} \cdot \bm{\nabla}}\Upsilon
                + \nu\p{\nabla^2 \Upsilon + \p{
                    \bm{\nabla} \Upsilon} \cdot \p{\bm{\nabla}\Upsilon}
                    - \frac{2}{H}\partial_z \Upsilon
                    + \frac{1 - e^{-\Upsilon}}{H^2}}\bigg],\\
        \pd{u_x}{t} + \pd{\varpi'}{x} + gH\pd{\Upsilon}{x} ={}&
            -\Gamma(z) u_x
            + \Gamma_{NL}\bigg[\nu \nabla^2 u_x
            - u_x \nu\p{\nabla^2 \Upsilon + \p{\bm{\nabla} \Upsilon} \cdot
                \p{\bm{\nabla}\Upsilon} - \frac{2}{H}\partial_z \Upsilon
                + \frac{1 - e^{-\Upsilon}}{H^2}}\nonumber\\
            &+ 2\nu \p{\p{\p{\bm{\nabla}\Upsilon} \cdot \bm{\nabla}}u_x
                - \frac{1}{H}\partial_z u_x}
                - \p{\bm{u} \cdot \bm{\nabla}}u_x
                - \varpi' \pd{\Upsilon}{x}\bigg],\\
        \pd{u_z}{t} + \pd{\varpi'}{z} + gH\pd{\Upsilon}{z} - \frac{\varpi'}{H}
            ={}& -\Gamma(z) u_z
            +\Gamma_{NL}\bigg[\nu \nabla^2 u_z
            - u_z \nu\p{\nabla^2 \Upsilon + \p{\bm{\nabla} \Upsilon} \cdot
                \p{\bm{\nabla}\Upsilon} - \frac{2}{H}\partial_z \Upsilon
                + \frac{1 - e^{-\Upsilon}}{H^2}}\nonumber\\
            &+ 2\nu \p{\p{\p{\bm{\nabla}\Upsilon} \cdot \bm{\nabla}}u_z -
                \frac{1}{H}\partial_z u_{z}}
            - \p{\bm{u} \cdot \bm{\nabla}}u_z
            - \varpi' \pd{\Upsilon}{z}\bigg].
    \end{align}
\end{subequations}
\label{lastpage} 
\end{document}